\documentclass[useAMS,usenatbib,breaklinks=true]{mnras}
\usepackage{graphicx}
\usepackage{times}
\usepackage{mn2e-breakabs}
\usepackage{amssymb}
\usepackage{amsmath}
\usepackage[dvipsnames]{xcolor}
\usepackage{comment}

\usepackage[normalem]{ulem} 

\newcommand{\com}[1]{\textcolor{black}{#1}}

\title[BAO in the projected cross-correlation function]{Baryon Acoustic Oscillations in the projected cross-correlation function between the eBOSS DR16 quasars and photometric galaxies from the DESI Legacy Imaging Surveys}

\author[P. Zarrouk et al.]{Pauline Zarrouk$^{1}$\thanks{E-mail: pauline.s.zarrouk@durham.ac.uk},
Mehdi Rezaie$^{2}$,
Anand Raichoor$^{3}$,
Ashley J. Ross$^{4}$,
Shadab Alam$^{5}$,
\newauthor
Robert Blum$^{6}$,
David Brookes$^{7}$,
Chia-Hsun Chuang$^{8}$,
Shaun Cole$^{1}$,
Kyle S. Dawson$^{9}$,
\newauthor
Daniel J. Eisenstein$^{10}$,
Robert Kehoe$^{11}$,
Martin Landriau$^{12}$,
John Moustakas$^{13}$,
Adam D. 
\newauthor
Myers$^{14}$,
Peder Norberg$^{1,15}$,
Will J. Percival$^{16,17,18}$,
Francisco Prada$^{18}$,
Michael Schubnell$^{20}$,
\newauthor
Hee-Jong Seo$^{2}$,
Gregory Tarl\'e$^{20}$,
Cheng Zhao$^{21}$
\\
\scriptsize $^{1}$ Institute for Computational Cosmology, Department of Physics, Durham University, South Road, Durham DH1 3LE, UK\\
\scriptsize $^{2}$ Department of Physics and Astronomy, Ohio University, 251B Clippinger Labs, Athens, OH 45701, USA\vspace*{-2pt} \\
\scriptsize $^{3}$ Institute of Physics, Laboratory of Astrophysics, \'Ecole Polytechnique F\'ed\'erale de Lausanne (EPFL), Observatoire de Sauverny, 1290 Versoix, Switzerland\vspace*{-2pt} \\ 
\scriptsize $^{4}$ Center for Cosmology and Astro-Particle Physics, Ohio State University, Columbus, Ohio, USA\vspace*{-2pt} \\ 
\scriptsize $^{5}$ Institute for Astronomy, University of Edinburgh, Royal Observatory, Blackford Hill, Edinburgh EH9 3HJ, UK \\
\scriptsize $^{6}$ National Optical Astronomy Observatory, 950 N. Cherry Ave., Tucson, AZ 85719, USA \\
\scriptsize $^{7}$ Department of Physics \& Astronomy, University College London, Gower Street, London, WC1E 6BT, UK \\
\scriptsize $^{8}$ Kavli Institute for Particle Astrophysics and Cosmology, Stanford University, 452 Lomita Mall, Stanford, CA 94305, USA
\scriptsize $^{9}$ Department of Physics and Astronomy, University of Utah, Salt Lake City, UT 84112, USA \\
\scriptsize $^{10}$ Center for Astrophysics, Harvard \& Smithsonian, 60 Garden St., MA 02138, USA \\
\scriptsize $^{11}$ Department of Physics, Southern Methodist University, 3215 Daniel Ave., Dallas, TX, 75205, USA \\
\scriptsize $^{12}$ Lawrence Berkeley National Laboratory, 1 Cyclotron Road, Berkeley, CA 94720, USA \\
\scriptsize $^{13}$ Department of Physics and Astronomy, Siena College, 515 Loudon Road, Loudonville, NY 12211, USA \\ 
\scriptsize $^{14}$ University of Wyoming, 1000 E. University Ave., Laramie, WY 82071, USA \\
\scriptsize $^{15}$ Centre for Extragalactic Astronomy, Department of Physics, University of Durham, South Road, Durham DH1 3LE, UK \\
\scriptsize $^{16}$ Waterloo Centre for Astrophysics, University of Waterloo, 200 University Ave W, Waterloo, ON N2L 3G1, Canada \\
\scriptsize $^{17}$ Department of Physics and Astronomy, University of Waterloo, 200 University Ave W, Waterloo, ON N2L 3G1, Canada \\
\scriptsize $^{18}$ Perimeter Institute for Theoretical Physics, 31 Caroline St. North, Waterloo, ON N2L 2Y5, Canada \\
\scriptsize $^{19}$ Instituto de Astrofısica de Andalucıa (CSIC), Glorieta de la Astronomıa, s/n, E-18008 Granada, Spain \\
\scriptsize $^{20}$ Department of Physics, University of Michigan, Ann Arbor, MI 48109, USA \\
\scriptsize $^{21}$ Institute of Physics, Laboratory of Astrophysics, \'Ecole Polytechnique F\'ed\'erale de Lausanne (EPFL), Observatoire de Sauverny, CH-1290 Versoix, Switzerland
}
\date{Accepted XXX. Received YYY; in original form ZZZ}

\pubyear{2020}

\begin{document}
\label{firstpage}
\pagerange{\pageref{firstpage}--\pageref{lastpage}}
\maketitle

\begin{abstract}
We search for the Baryon Acoustic Oscillations in the projected cross-correlation function binned into transverse comoving radius between the SDSS-IV DR16 eBOSS quasars and a dense photometric sample of galaxies selected from the DESI Legacy Imaging Surveys. We estimate the density of the photometric sample of galaxies in this redshift range to be about 2900 deg$^{-2}$, which is deeper than the official DESI ELG selection, and the density of the spectroscopic sample is about 20 deg$^{-2}$. In order to mitigate the systematics related to the use of different imaging surveys close to the detection limit, we use a neural network approach that accounts for complex dependencies between the imaging attributes and the observed galaxy density. We find that we are limited by the depth of the imaging surveys which affects the density and purity of the photometric sample and its overlap in redshift with the quasar sample, which thus affects the performance of the method. When cross-correlating the photometric galaxies with quasars in $0.6 \leq z \leq 1.2$, the cross-correlation function can provide better constraints on the comoving angular distance, $D_{\rm M}$ (6\% precision) compared to the constraint on the spherically-averaged distance $D_{\rm V}$ (9\% precision) obtained from the auto-correlation. Although not yet competitive, this technique will benefit from the arrival of deeper photometric data from upcoming surveys which will enable it to go beyond the current limitations we have identified in this work. 

\end{abstract}



\section{Introduction}
\label{sec:intro}

The Baryon Acoustic Oscillations (BAO) feature~\citep{Eisenstein+05,Cole+05} in the clustering of galaxies left by the baryon-photon plasma which propagated as sound waves until decoupling in the early universe has emerged as a very robust way of measuring cosmic distances across time. Indeed, the BAO measurement in samples of galaxies at different redshifts is a powerful geometrical test to probe the expansion history of the universe in a complementary way to CMB anisotropies~\citep{Planck18} and to the Hubble diagram with Type 1a Supernovae for the local universe, such as the recent HST program SH$_{0}$ES~\citep{Riess+18,Riess+19} and the Carnergie-Chicago Hubble program~\citep{Freedman+19}.
These various cosmological data sets have shown increasing evidence that the cosmic expansion has been accelerating for 6 billion years. However, the mechanism which is driving such an acceleration is one of the biggest mysteries in cosmology. It is commonly known as 'Dark Energy' where, in the standard cosmological model, we introduce the cosmological constant $\Lambda$ which is associated with vacuum energy to account for this late-time acceleration. The current observations could also be explained by more complex dark energy models with time-dependent properties or even a modification of General Relativity at cosmological scales. Ongoing and planned cosmic surveys have been designed to distinguish between these possibilities and more generally to test the standard model of cosmology, $\Lambda$CDM.

Dense spectroscopic samples of large-scale structure tracers are required to measure the BAO feature accurately, but it becomes observationally expensive especially at high redshifts ($z \geq 1$) where galaxies are fainter and less abundant. Bright galaxies such as the BOSS luminous red galaxies (BOSS LRG \cite{boss-DR12}) have been extensively probed by spectroscopic surveys to reconstruct the map of the large-scale structures of the universe up to $z < 0.6$.
The SDSS-IV eBOSS program~\citep{eboss} undertook a survey of emission line galaxies and quasars to probe the unexplored intermediate redshift range ($0.6 < z < 2.2$) and set the scene for the advent of Stage IV dark energy experiments including the Dark Energy Spectroscopic Instrument at the Kitt Peak Observatory~\citep{Desi16a,Desi16b}, the space-mission \textit{Euclid}~\citep{Euclid13} and the Legacy Survey of Space and Time at the Vera-Rubin Observatory~\citep{LSST12}.
Current and upcoming galaxy surveys are also providing larger and larger samples of photometric data. Attempts to measure the BAO feature in the correlation function of photometric galaxy samples~\citep{Padmanabhan+07,Ross+15} continue improving with the ongoing generation of surveys like Hyper-Subprime Camera HSC~\citep{HSC}, the Dark Energy Survey~\citep{DES17} and the Kilo-Degree Survey KIDS~\citep{KIDS}.
However, the photometric BAO analysis is based on the angular correlation function of galaxies which loses the clustering information along the line-of-sight direction due to projection. It thus mixes different physical scales which leads to a smearing of the BAO feature and to a loss of precision in the cosmological constraints.

On the one hand, spectroscopic data provide accurate redshift measurements but are limited by the statistics at high redshift ($z > 1$). On the other hand, samples of photometric data are much larger but with less precise information in the radial direction. Therefore, exploiting the cross-correlation between both types of data has been more and more used, mainly in order to characterise the properties of the photometric dataset. For instance, \citet{Padmanabhan+09} studied the small-scale clustering of a sample of photometrically selected LRG using a spectroscopic sample of quasars in the range $0.2 < z < 0.6$. Methods to infer the redshift distribution of a photometric sample by measuring the amplitude of cross-correlation with spectroscopic samples have been developed~\citep[for a review, see][]{Newman08} and this technique has been recently applied to the photometric sample of DESI LRGs and was able to recover the expected redshift distribution~\citep{Kitanidis+19}.
Another application of such kind of cross-correlation was proposed by~\citet{Patej+18} with the goal of improving the BAO measurement in a sparse spectroscopic sample by exploiting the cross-correlation with a denser photometric sample.
\citet{Nishizawa+13} already showed that the measurement of the correlation function as a function of transverse comoving radius rather than angular separation preserves the BAO scale inherent in the large-scale structures. Then, \citet{Patej+18} derived the analytical prediction in configuration space and they presented a proof-of-concept that the BAO scale could be measured in the cross-correlation function binned by transverse comoving radius.

In this paper, we develop and apply this method to measure the BAO feature in the projected cross-correlation of eBOSS DR16 quasars and a photometric sample of galaxies selected from the DESI Legacy Imaging Surveys~\citep{Dey+19}. We perform two analyses in parallel, one from the auto-correlation function of the eBOSS DR16 quasars and one from the projected cross-correlation function, both under the same fitting conditions. By providing a proper comparison between the two statistics, our goal is to highlight the potential benefit of using the cross-correlation between photometry and spectroscopy for BAO measurement.
The paper is structured as follows. Section~\ref{sec:proj-corr} presents the analytical prediction for the projected cross-correlation function. Then, we present the spectroscopic and photometric data sets in Section~\ref{sec:data} and the methodology for this analysis in Section~\ref{sec:method}. Eventually, the measurements and results are shown and discussed in Section~\ref{sec:results} followed by the conclusion in Section~\ref{sec:concl}.

\section{BAO in the projected cross-correlation function}
\label{sec:proj-corr}

\subsection{Transverse comoving separation}
Instead of considering angular separation between a spectroscopic object and the surrounding photometric galaxies, the key idea is to consider the cross-correlation binned by transverse comoving separation where photometric galaxies are assumed to be at the same redshift as the spectroscopic object they are correlated with. The transverse separation $\mathbf{R}$ is thus defined in terms of observed angular positions and the comoving angular diameter distance evaluated at the spectroscopic redshift $D_{\rm M}(z_{\rm s})$:
\begin{equation}
    \mathbf{R} = D_{\rm M}(z_{\rm s}) \arccos(\boldsymbol{\gamma}_{\rm s} \cdot \boldsymbol{\gamma}_{\rm p}) = D_{\rm M}(z_{\rm s}) |\boldsymbol{\theta}_{\rm s} - \boldsymbol{\theta}_{\rm p}|
\label{eq:trans-radius}
\end{equation}
where the unit vector on the celestial sphere $\boldsymbol{\gamma}$ is defined by $\boldsymbol{\gamma} = (\sin \theta \cos \phi, \sin \theta \sin \phi, \cos \theta)$ \com{and $\boldsymbol{\theta}_{\rm s}$ (resp. $\boldsymbol{\theta}_{\rm p}$), is the angular separation between the spectroscopic quasar (resp. the photometric galaxy) and the line-of-sight direction.} We assume the sky is locally flat over separations between correlated galaxies. 

The conversion from angular position to transverse comoving separation assumes a fiducial cosmological model and also in the  next section, we will assume a linear matter power spectrum. We use a flat $\Lambda$CDM cosmological model with the following parameters:
\begin{equation}
    \centering
    \begin{split}
    h &= 0.676, \,\, \Omega_{m} = 0.31, \,\,  \Omega_{\Lambda} = 0.69,\\
    \Omega_{b}h^{2} &= 0.022, \,\, m_{\nu}~= 0.06~eV \,\, \sigma_{8} = 0.80.\qquad
    \end{split}
    \label{eq:fiducial-cosmology}
\end{equation}
where the subscripts $m$, $b$ and $\nu$ stand for matter, baryon and neutrino, respectively, and $h$ is the standard dimensionless Hubble parameter. These choices match the fiducial cosmology adopted for the BAO analysis of the eBOSS quasars sample DR14~\citep{DR14-bao} and DR16~\citep{Hou+20,Neveux+20}.

\subsection{Analytical projected cross-correlation function}
\citet{Patej+18} detailed the mathematical formalism associated with the projected cross-correlation function $w_{\theta}(\mathbf{R})$. In this section, we just recall the most important steps in obtaining an analytical prediction for $w_{\theta}(\mathbf{R})$. The spectroscopic sample is characterised by an over-density field $\delta_{\rm s}(\mathbf{r})$ and the photometric sample by a projected over-density field $\Delta_{\rm p}(\boldsymbol{\theta})$ defined as:
\begin{equation}
    \Delta_{\rm p}(\boldsymbol{\theta}) = \frac{\int r^{2} n_{\rm p}(r) \delta_{\rm p}(\mathbf{r}) dr} {\int r^{2} n_{\rm p}(r) dr}
\end{equation}
where $n_{\rm p}(r)$ is the number density of photometric galaxies and $\delta_{\rm p}(\mathbf{r})$ the over-density field associated with the photometric sample.

The projected cross-correlation function $w_{\theta}(\mathbf{R})$ is then:
\begin{equation}
    w_{\theta}(\mathbf{R}) = \langle \delta_{\rm s}(\mathbf{r}) \Delta_{\rm p}(\boldsymbol{\theta}+\mathbf{R}/r) \rangle
\end{equation}
where the transverse comoving separation $\mathbf{R}$ is defined by equation~\ref{eq:trans-radius} and $\mathbf{r} = (r,\boldsymbol{\theta})$.

Using the flat-sky approximation, \citet{Patej+18} derived a detailed analytical expression in agreement with \citet{Padmanabhan+09} who first showed that the projected correlation function binned by physical transverse separation $w_{\theta}(\mathbf{R})$ can be related to the standard projected correlation $w_{\rm p}(\mathbf{R})$~\citep{Davis+83} by:
\begin{equation}
    w_{\theta}(\mathbf{R}) = \langle f(b_{s},n_{s},b_{p},n_{p}) \rangle w_{\rm p}(\mathbf{R})
    \label{eq:wR-wp}
\end{equation}
where $\langle f(\chi) \rangle$ scales the amplitude of the standard projected cross-correlation function according to the overlap in redshift between the photometric galaxies and the spectroscopic quasars they are correlated with. This factor is defined in~\citet{Patej+18} as follows:
\begin{equation}
    \langle f(b_{s},n_{s},b_{p},n_{p}) \rangle  = \frac{b_{\rm s}b_{\rm p}}{2\pi} \frac{\int dr \, r^{2} n_{\rm s}(r) W(r,\eta) n_{\rm p}(r)} {\int dr \, n_{\rm s}(r) W(r,\eta) \int dr' \, r'^{2} n_{\rm p}(r')}
    \label{eqn:wR-factor}
\end{equation}
where $b_{\rm s}$ is the linear bias of the spectroscopic sample, $b_{\rm p}$ the linear bias of the photometric sample, $n_{\rm s}$ is is the redshift-dependent mean number density of the spectroscopic sample, $n_{\rm p}$ the redshift-dependent mean number density of the photometric sample and $W(r,\eta)$ is a weighting function which accounts for selection effects and can depend on $r$ and other variables which are collectively referred to as $\eta$.
However, so far in the clustering analyses, the weights that have been derived do not depend on $r$ or $\eta$ and we generally assume that the linear biases are scale-independent, therefore equation~\ref{eqn:wR-factor} can be seen as a normalisation factor which will be fitted to the data. If the two redshift distributions were overlapping perfectly, the second ratio would be equal to 1 and the normalisation factor would simply be $b_{\rm s}b_{\rm p}/2\pi$. 

The standard projected correlation function $w_{\rm p}(\mathbf{R})$ can be calculated as the line-of-sight integral of the two-dimensional power spectrum defined as:
\begin{equation}
    w_{\rm p}(\mathbf{R}) = \frac{1}{2\pi} \int d^{2}k_{\perp} P(k_{\perp}) e^{i \mathbf{k}_{\perp} \cdot \mathbf{R}}
\end{equation}
where $k = \sqrt{k^{2}_{\parallel} + k^{2}_{\perp}} \approx k_{\perp}$ assuming that the range of the line-of-sight distance projected over is much larger than the BAO scale.

However, the expectation value derived for $w_{\theta}(\mathbf{R})$ in equation~\ref{eq:wR-wp} only holds for a single separation $\mathbf{R}$ whereas the data are binned. So to account for this binning, one can exploit that the clustering depends only on $r$ due to isotropy. Therefore, by integrating over circular annuli with bounds $R_{1}$ and $R_{2}$, we can obtain the following binned correlation function:
\begin{equation}
    w(R_{1},R_{2}) = 2 \int_{R_{1}}^{R_{2}} \frac{R dR}{(R_{2}^{2}-R_{1}^{2})} \int \frac{d\phi}{2\pi}w_{\theta}(\mathbf{R})
\end{equation}
which then leads to equation (80) of \citet{Patej+18}:
\begin{equation}
    \begin{aligned}
    w(R_{1},R_{2}) = \frac{b_{\rm s}b_{\rm p}} {2\pi^{2}(R_{2}^{2}-R_{1}^{2})} \frac{\int dr \, r^{2} n_{\rm s}(r) W(r,\eta) n_{\rm p}(r)} {\int dr \, n_{\rm s}(r) W(r,\eta) \int dr' \, r'^{2} n_{\rm p}(r')} \\
    \int d^{2}k_{\perp} P(k_{\perp}) \frac{R_{2}J_{1}(k_{\perp}R_{2}) - R_{1}J_{1}(k_{\perp}R_{1})} {k_{\perp}}
    \end{aligned}
\label{eq:binned-wR}
\end{equation}
This is the analytical expression we will use to fit to the data in Section~\ref{sec:bao-cross}.

\subsection{Implicit assumptions}

\begin{figure}
    \includegraphics[width=\columnwidth]{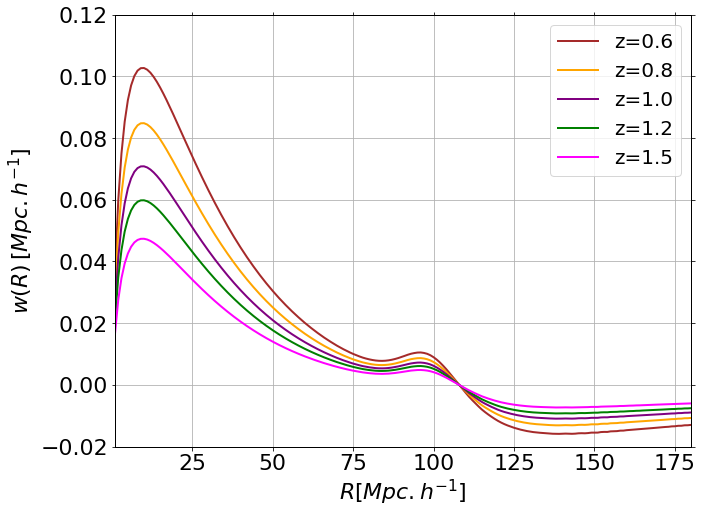}
    \caption{Evolution with redshift of the template for \com{the matter projected correlation function} following equation~\ref{eq:binned-wR} and with a bin width of 1~$h^{-1}$Mpc.}
    \label{fig:wR-zevol} 
\end{figure}

The above derivation implicitly assumes that $w_{\rm p}(\mathbf{R})$
does not vary across the redshift range of interest. Of course, it is not true and Fig.~\ref{fig:wR-zevol} shows the redshift evolution of the \com{matter} projected cross-correlation function using equation~\ref{eq:binned-wR} for a bin width of 1~$h^{-1}$Mpc. However, in the context of BAO measurements we argue that the position of the BAO peak is not affected and any modification of the shape of the projected cross-correlation can be accounted for with the bias and broad-band parameters. Nevertheless, in Section~\ref{sec:effective-z} we propose three assumptions for the redshift distribution of the photometric sample that can affect the effective redshift definition and in Section~\ref{sec:bao-cross} we will show the impact on the BAO measurements.

Another assumption is related to the input matter power spectrum $P(k_{\perp})$ which enters equation~\ref{eq:binned-wR}. We generate it using CAMB~\citep{camb} assuming a linear prediction. \com{and We have checked that including non-linear corrections from the Halofit model~\citep{Halofit03,Halofit12} in CAMB has a marginal effect on small scales ($R \leq 20 \, h^{-1}$Mpc). Moreover, the projected clustering is less affected by redshift-space distortions~\citep{Ross+11} and we do not use photometric redshifts to locate the position of the galaxies along the line of sight. So we expect the RSD correction to be negligible. However, \citet{EisensteinSeoWhite07} showed that non-linearities can also lead to a broadening of the BAO peak, however this evolution is sufficiently slow that it can be accounted for by adding a damping term in the BAO template. Eventually, we can also have a scale-dependent bias that becomes important at small scales, however in a BAO-analysis it is usually accounted for by the broad-band parameters. We refer to Section~\ref{sec:baofit} where we present the BAO fitting procedure and introduce the damping term and the broad-band parameters.}


\section{Data}
\label{sec:data}

\subsection{Spectroscopic dataset}
\label{sec:spectro-data}

As highlighted in~\citet{Patej+18}, the cross-correlation method described above is expected to reduce shot noise and then to improve the BAO measurement in a sparse spectroscopic sample. Actually, the SDSS-IV eBOSS quasars represent a sparse spectroscopic sample with $nP_0 = 0.12 < 1$ where $n \simeq 10^{-4} \, h^{-3}$Mpc$^{-3}$ is the quasar density (which is about an order of magnitude lower than for galaxies) and $P_0=6000 \, h^{-3}$Mpc$^3$ is the typical amplitude of the quasar power spectrum at the BAO scale.

\begin{figure}
    \includegraphics[width=\columnwidth]{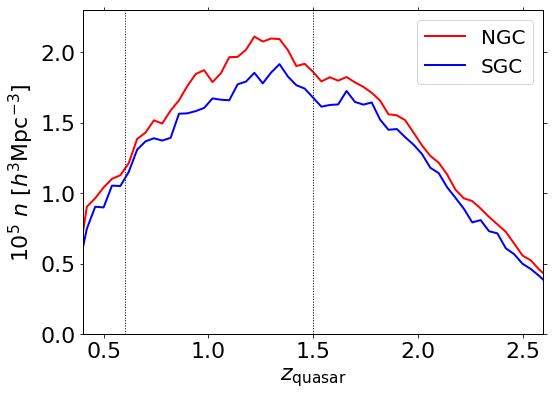}
    \caption{Redshift distribution of the spectroscopic eBOSS quasars. We keep quasars in $0.6 \leq z \leq 1.5$ only as indicated by the dotted lines.}
    \label{fig:spectro-z} 
\end{figure}

Our analysis uses the CORE spectroscopic sample of quasars obtained by the SDSS-IV eBOSS program~\citep{eboss} using the 2.5m Sloan Foundation telescope \citep{Gunn+06} at the Apache Point Observatory with the same two-arm optical fibre-fed spectrographs as BOSS~\citep{Smee+13}. The CORE quasar target selection~\citep{Myers+15} is based on the SDSS-I-II-III optical imaging data in the \textit{ugriz}~\citep{Fukugita+96} photometric pass band and on the Wide-field Infrared Survey Explorer~\citep[\textit{WISE}][]{Wright+10}. Selection is performed using a likelihood-based routine called the ``Extreme Deconvolution" algorithm (XDQSO) in order to obtain a homogeneous quasar sample at $g < 21$. The algorithm has been improved for eBOSS with XDQSOz~\citep{Bovy+12} in order that it can be applied to any redshift range.
The spectroscopic sample used in this analysis is obtained with the same methodology as for the large-scale structure DR16 eBOSS quasar catalogue~\citep{Ross+20} but instead of considering quasars between $0.8 \leq z \leq 2.2$, we use quasars between $0.6 \leq z \leq 1.5$. Including quasars at $z < 0.8$ and removing those at $z > 1.5$ allows us to increase the overlap in redshift with the sample of photometric galaxies. There are very few quasars below $z=0.6$ and few photometric galaxies above $z=1.5$. The photometric sample is described in the next section.
The redshift distribution of the eBOSS DR16 quasar catalogue~\citep{Lyke+20} is shown in Fig.~\ref{fig:spectro-z} in blue for the NGC and red for the SGC. The dotted vertical lines delineate our range $0.6<z< 1.5$. 

\begin{figure}
    \includegraphics[width=\columnwidth]{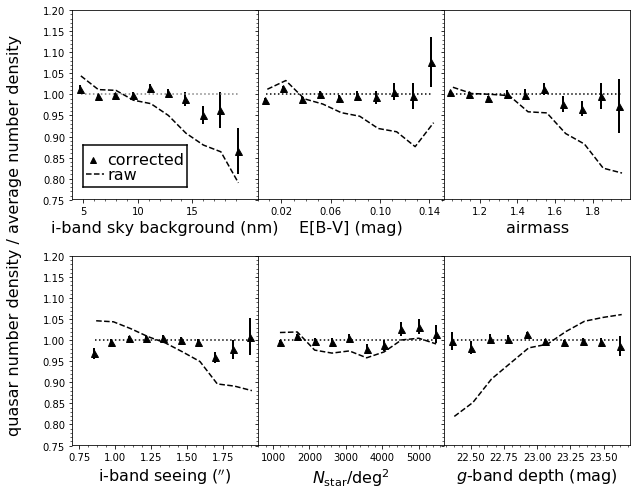}
    \caption{Quasar target density variation with imaging systematics before (raw data as dotted lines) and after applying weights to correct for the variation with $g$-band depth and Galactic extinction (E[B-V]). Here, we consider quasars in $0.6 \leq z \leq 1.5$.}
    \label{fig:spectro-syst} 
\end{figure}

As in~\citet{Ross+20}, we apply a cut in the completeness per sector where we restrict to sectors that have $C_{\rm eBOSS} > 0.5$ such that the completeness of the quasar sample is $\sim$ 97.7\%.
We also correct for any missing targets or spurious correlation in the target density by applying several weights to the data and the randoms following~\citet{Ross+20}:
\begin{itemize}
    \item Systematics weight $w_{\rm sys}$ are used to correct for inhomogeneities in the quasar density due to variations in the quality of the SDSS photometry. Such variations can lead to angular variations of the depth (5$\sigma$ detection in magnitude for a point-source object) which also depend on the airmass, seeing and Galactic extinction. We compute systematics weights for the NGC and SGC separately based on linear regression according to the dependence of the quasar density on the SDSS imaging depth in the $g$-band, Galactic extinction, the seeing, and the sky background. The procedure was identical to that described in \citet{Ross+20}, except for the $0.6 < z < 1.5$ redshift range. Fig.~\ref{fig:spectro-syst} displays the target density variation with the imaging systematics before applying the correction with depth and Galactic extinction (raw data) and after (corrected). We then correct for the remaining dependence on seeing and sky background.
    \item Close-pairs (fibre-collisions) weight $w_{\rm cp}$ are used to account for the missing targets in a collision group due to the finite size of a spectroscopic fibre which prevents observing two quasars within a radius of 62$^{\prime\prime}$. We compute weights which are assigned and equally distributed per collision group.
    \item Redshift failures weight $w_{\rm noz}$ are used to account for the missing targets due to invalid redshifts that we correct for by computing weights based on the spectrograph signal to noise in the $i$-band and the fibre ID.
    \item FKP weight $w_{\rm FKP}$~\citep{FKP} are used to minimise the variance of the measurement and which is defined by $w_{\rm FKP}=\left(1+P_0 n(z)\right)^{-1}$.
\end{itemize}
The total weight which is applied to the data and the random is defined by:
\begin{equation}
w_{\rm tot}=w_{\rm FKP}\cdot w_{\rm sys}\cdot w_{\rm cp}\cdot w_{\rm noz}
\label{eqn:weight_tot}
\end{equation}

\subsection{Photometric dataset}
\label{sec:photo-data}

In order to maximise the cross-correlation signal, we need to select photometric galaxies in the same footprint but also in the same redshift range as the spectroscopic sample. We also need to reach a sampling of galaxies dense enough such that $n_{\rm p}P_{\rm p} > 1$ where $n_{\rm p}$ is the density of the photometric galaxies and $P_{\rm p}$ is the typical value of the power spectrum of the photometric sample at the BAO scale.
Given our choice of spectroscopic sample, we want to select photometric galaxies in the redshift range $0.6 < z < 1.5$, where star-forming galaxies are ideal candidates as they are abundant at these redshifts with strong emission lines~\citep[e.g.,][]{Madau+14}.
Such kind of Emission Line Galaxies (ELG) have been already observed by the SDSS-IV eBOSS program which confirms that optical colour selection techniques can be used to optimally select ELGs in $0.6 < z < 1.1$~\citep{Raichoor+17}. The DESI instrument~\citep{Desi16b} has been designed to resolve the [OII] doublet over the redshift range $0.6 < z < 1.6$ such that the ELGs constitute the largest sample of objects that DESI will observe with 28 million ELGs over 14,000 deg$^{2}$.
We select the photometric sample using the DR8 release~\footnote{\url{http://legacysurvey.org/dr8/description/}} of the DESI Legacy Imaging Surveys~\citep{Dey+19} which consist of:
\begin{itemize}
    \item the Dark Energy Camera Legacy Survey (DECaLS) provides imaging over 2/3 of the DESI footprint covering both the Northern and Southern Galactic caps (NGC and SGC) at Dec $\leq 32$ deg in $g$, $r$ and $z$ bands. It also includes the DES imaging where available.
    \item the Beijing-Arizona Sky Survey (BASS) observes 5500 deg$^{2}$ in the NGC footprint at Dec $\geq 32 $ deg in two optical bands ($g$ and $r$). Its coverage includes $500$ deg$^{2}$ of overlap  with DECaLS in order to investigate any systematic biases in the target selection.
    \item the Mayall $z$-band Legacy Survey (MzLS) observes 5500 deg$^{2}$ in the NGC footprint at Dec $\geq 32$ deg in the $z$ band.
\end{itemize}

\begin{figure}
    \centering
    \includegraphics[width=\columnwidth]{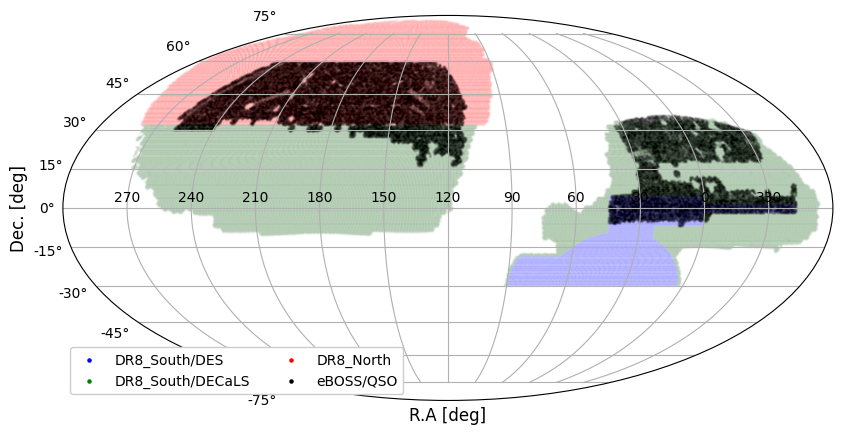}
    \caption{Footprint of the DESI Legacy Imaging Surveys with BASS/MzLS (red), DECaLS (red) and DES (blue) and of the DR16 eBOSS/QSO footprint (black).}
    \vskip -0.2cm
    \label{fig:footprint-surveys}
\end{figure}

Altogether, they provide photometric data in three optical/near-infrared bands ($g$,$r$ and $z$) over more than 14,000 deg$^{2}$ with a $5\sigma$ galaxy depths of $g=24.4$, $r=23.8$ and $z=23.0$. Note that the separation between BASS/MzLS and DECaLS is at Dec $= 32.375$ deg and that this declination does not reflect the limits of the imaging but, rather, is imposed by the \textsc{desitarget} code~\footnote{\url{https://github.com/desihub/desitarget/blob/master/py/desitarget/io.py\#L95}}. The photometric catalogue also includes two mid-infrared bands observed by the \textit{WISE} satellite~\citep{Wright+10}.
All the Legacy Imaging Surveys catalogues are built using \textsc{Tractor}~\footnote{\url{https://github.com/dstndstn/tractor}}~\citep{Lang+16} which is a forward-modelling algorithm to perform source extraction on pixel-level data.
The footprint of the different DESI Legacy Imaging Surveys with BASS/MzLS, DECaLS and DES together with the eBOSS footprint are shown in Fig.~\ref{fig:footprint-surveys}. 

\subsubsection{Photometric selection}
Our target selection is based on three criteria: i) clean photometry (masking around bright stars/objects and removing defective pixels), ii) high target density, iii) galaxies in the desired redshift range. The corresponding cuts are detailed in Table~\ref{tab:photo-selection} and we modify the \textsc{desitarget} code~\footnote{\url{https://github.com/desihub/desitarget/tree/master/py/desitarget}} to implement them.

Our cuts are done on magnitudes corrected for Galactic extinction using the maps of~\citep{SFD98}.
The extinction coefficients for the DECam filters are computed using $airmass = 1.3$ for a source with a $7000$~K thermal spectrum as done in~\citep{SchlaflyFinkbeiner11}. These coefficients are ${\rm A / E(B-V)} = 3.995$, $3.214$, $2.165$, $1.592$, $1.211$, $1.064$ for the g, r, z, W1 and W2 bands respectively. Galactic extinction coefficients for BASS and MzLS are also calculated as if they are on the DECam filter system. 
\textsc{Tractor} outputs also provide the number of observations $n_{\rm obs}$ in the three bands where we impose at least one observation in each band ($n_{\rm obs, g/r/z} > 0$), a masking around GAIA DR2 stars~\citep{Gaia-dr2} and corrections for instrumental effects that we call `pixel masking'. These corrections enable us to track the pixels that have been compromised due to bad quality, saturation, cosmic rays, bleed trails, transients, edges and outliers. They are compiled in the ALLMASK bitmask~\footnote{\url{http://legacysurvey.org/dr8/bitmasks/\#maskbits}} and we remove those bad pixels for each band. Our pixel masking also includes pixels in the vicinity of bright stars, large galaxies\footnote{\url{https://github.com/moustakas/LSLGA}} and globular clusters.

\begin{table*}
\centering
\caption{The ELG target selection in this study using the DESI Imaging Legacy Surveys in the NGC and SGC.}
\label{tab:photo-selection}
\begin{tabular}{|c|c|c|}
 Criterion & DESI BASS/MzLS & DESI DECaLS \\
\hline
Clean photometry  & \multicolumn{2}{c}{$n_{\rm obs, g/r/z} > 0$, GAIA stars masking and pixel masking} \\
Magnitude range   & $20 < g < 23.6$ & $20 < g < 23.5$  \\
Redshift range    & \multicolumn{2}{c}{$1.15(r-z) - 1.5 < g-r < 1.15(r-z) - 0.15$} \\
                  & \multicolumn{2}{c}{$-1.2(r-z) + 1.0 < g-r < -1.2(r-z) + 2.8 $} \\
\hline
\end{tabular}
\end{table*}

\begin{figure}
	\centering
    \includegraphics[width=\columnwidth]{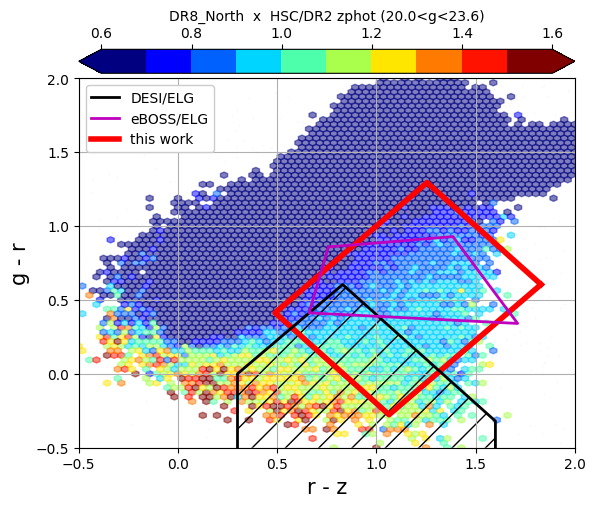}
    \includegraphics[width=\columnwidth]{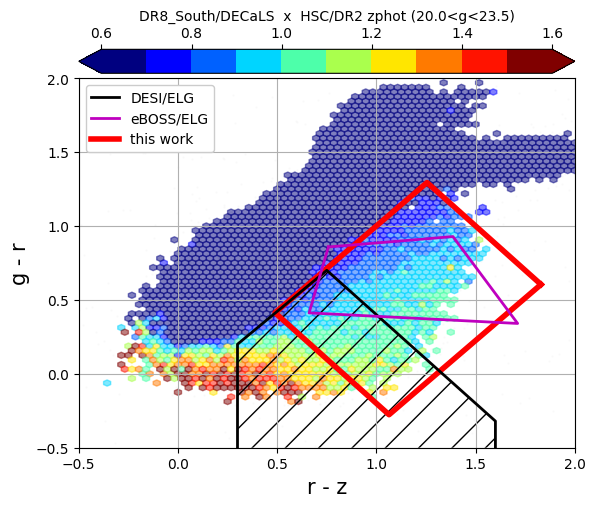}
    \vskip -0.2cm
    \caption{Colour-colour diagram showing the selection used in this analysis (red), the DESI ELG main selection (black) and the eBOSS ELG selection (magenta). The colour-codding represents photometric redshifts from HSC-PDR2. Top: in BASS/MzLS, bottom: in DECaLS.}
    \label{fig:grz-elg} 
\end{figure}

Fig.~\ref{fig:grz-elg} displays the $g-r$ vs $r-z$ colour-colour diagram for galaxies in our $g$-band magnitude range with our selection (red box), the DESI ELG selection (black box) and the SDSS-IV eBOSS ELG selection (magenta box). The colour-coding shows the photometric redshifts of the galaxies after matching the ELG targets with HSC-PDR2~\citep{HSC-pdr2}.
The DESI ELG target selection does not provide enough targets for this study, with a mean target density of 2400 deg$^{-2}$ while we would like a mean density of $\sim$3000 deg$^{-2}$ to have $n_{\rm p}P_{\rm p} \sim 5$ where we choose $P_0 = 4 000 \, h^{-3}$ Mpc$^3$ for ELGs.
To reach a higher redshift density, we select less star-forming objects (i.e with redder colours) given that we are not constrained by desiring a minimal [OII] flux as spectroscopic targets are. 

Because we are only interested in the overlapping footprint with eBOSS, we apply the eBOSS geometry to the photometric sample including four veto masks for bad fields, bright objects, a centre post mask which removes the areas at the centre of the plates where no targets can be observed and a collision priority mask which removes the areas where higher priority targets prevent any fibre being assigned to a quasar target.

\begin{figure}
	\centering
    \includegraphics[width=\columnwidth]{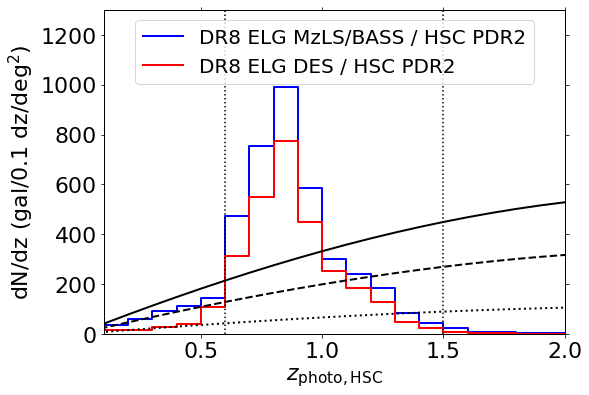}
    \caption{Photometric redshift distribution of the matched galaxies between our photometric sample and HSC PDR2. The blue curve shows the redshift distribution of the matched objects in the BASS/MzLS region and the red curve in the DES region. The dotted, dashed and solid curves represent respectively $n_{\rm p}P_{\rm p}=1$, $n_{\rm p}P_{\rm p}=3$ and $n_{\rm p}P_{\rm p}=5$.}
    \label{fig:photo-z} 
\end{figure}

In order to validate our selection, we match our catalogue after applying the colour-cuts and the eBOSS geometry with HSC-PDR2 for the NGC and SGC separately. Fig.~\ref{fig:photo-z} shows the resulting redshift distribution for the matched objects in the NGC (blue) and SGC (red) using the HSC-PDR2 photometric redshifts~\citep{HSC-pdr2-photoz}. The black lines shows respectively the $n_{\rm p}P_{\rm p}=1,3,5$ surface density when evaluated at wave number $k=0.14 h$Mpc$^{-1}$ and orientation relative to the line of sight $\mu=0.6$. These surface densities assume a fiducial constant bias $b(z)D(z)=0.84$ from DEEP2 ELG data~\citep{Mostek+13} where $D(z)$ is the linear growth factor normalised by $D(z=0)=1$. We can see that the photometric redshift distribution is above $nP=5$ for $0.6 < z < 1.2$, meaning that in this redshift range we reach a high sampling with an approximate target density of 2900 deg$^{-2}$. For this reason, we decide to explore the performance of the cross-correlation technique in two redshift ranges: $0.6 \leq z \leq 1.2$ and $0.8 \leq z \leq 1.5$. In the latter, we can expect the cross-correlation technique to be less efficient due to a limitation in our capability of selecting a denser sample of galaxies at $z > 1.2$ as we are already pushing the selection to very faint objects ($g_{\rm faint \, end}=23.5-23.6$), at the limit of detection given the imaging surveys we use ($g_{5\sigma}=24$).

\subsubsection{Imaging systematics and mitigation technique}
As for the spectroscopic sample, the number density of ELGs in the photometric sample suffers from observational systematics that arise because of inhomogeneities in the imaging. In order to minimise the impact of these inhomogeneities on our estimate of the true galaxy overdensity field, we apply weights to the photometrically-selected galaxies. 
So far, previous studies using the large-scale structure catalogues such as for BOSS DR12 galaxies~\citep{Reid+16} and for eBOSS DR16 tracers~\citep{Ross+20}, were based on multi-variate regression techniques to model the dependency between the imaging systematics and the observed target density by usually assuming a linear or quadratic relation. However, for strong contamination like the one close to the Galactic plane, this assumption may be no longer valid. Moreover, the correlations between the Legacy Survey imaging bands are more complex than the ones in SDSS and may not be fully captured by a linear model. Because we are selecting very faint objects at the limit of the survey depth, we also expect the ELG selection to be prone to more fluctuations. For all these reasons, recent progress has been made to develop more advanced systematics mitigation technique, such as the one in~\citet{Rezaie+19} based on artificial neural networks (NN) that has been applied to the DECaLS DR7 data with the eBOSS ELG selection. \com{The approach implements a 5-fold partitioning of the data that allows permutation of training, validation, and testing over the entire footprint thereby without a need for multiple realizations of the sky. The methodology presents a multi-layer neural network with non-linear activation function on the hidden layers that provides a non-linear mapping between the input imaging maps and observed density of ELGs. The network parameters are trained using gradient descent with batches of pixels and minimizing the sum of the residual squared error between the observed density of ELGs and the output of the network, plus an additional L2 regularization term (i.e., proportional to the sum of parameters squared) to suppress over-fitting. The hyperparameters include the number of hidden layers, regularization scale, and batch size which are tuned by applying the trained network on the validation set. Ultimately, the network with the best set of hyper-parameters is applied to the test set. Assuming there is no correlation between the cosmological signal and input imaging maps, the output of the regression, called the  \textit{selection mask},represents solely the systematic effects in the observed density and therefore its inverse can be applied as a weight to galaxies to mitigate the imaging systematics. \cite{Rezaie+19} illustrated that the non-linear neural network regression reduces the excess clustering on the largest scales more effectively than the conventional, linear regression.}

We apply this technique to the photometric sample used in this analysis and for comparison, we also derive weights based on the standard multivariate linear regression assuming both linear and quadratic terms. In what follows, we summarise the main steps to derive weights based on the NN and a detailed description of the methodology can be found in~\citet{Rezaie+19}:
\begin{enumerate}
\item We first produce the HEALPIX maps~\citep{Gorski+05} by splitting the sky into equal-area pixels for the imaging attributes that we consider as potential sources of systematics based on the DR8 ccds-annotated file~\footnote{\url{http://www.legacysurvey.org/dr8/files/\#ccds-annotated-camera-dr8-fits-gz}} using the \textsc{validationtests} pipeline~\footnote{\url{https://github.com/legacysurvey/legacypipe/tree/master/validationtests}}\com{, a modified implementation of \textsc{QuickSip}~\footnote{\url{https://github.com/ixkael/QuickSip}}~\citep{Leistedt15}}. These maps have a resolution of 13.7 arcmin ($nside = 256$) and we use the following imaging quantities: Galactic extinction \citep{SFD98}, stellar density \citep{Gaia-dr2}, \com{and} hydrogen atom column density~\citep{Bekhti+16}, \com{as well as galaxy depth,} sky brightness, seeing\com{, airmass, exposure time, and Modified Julian Date in r, g, and z pass bands.} In total, we have 21 CCD-based maps.

\item We use the same modelling and setting parameters for the neural network, i.e. the number of hidden layers, type of non-linear activation function and numbers of neurons in each layer as in~\citet{Rezaie+19}. We also use 5 folds to train the parameters, tune the hyper-parameters and to estimate the performance of the method. The neural network takes as input the imaging attributes \com{as independent variables} and the galaxy density \com{as target variable}, then it provides as output an estimate of the selection mask (or contamination model) whose inverse corresponds to the photometric weights we can apply to the data.

\end{enumerate} 

\begin{figure}
	\centering
	\includegraphics[width=\columnwidth]{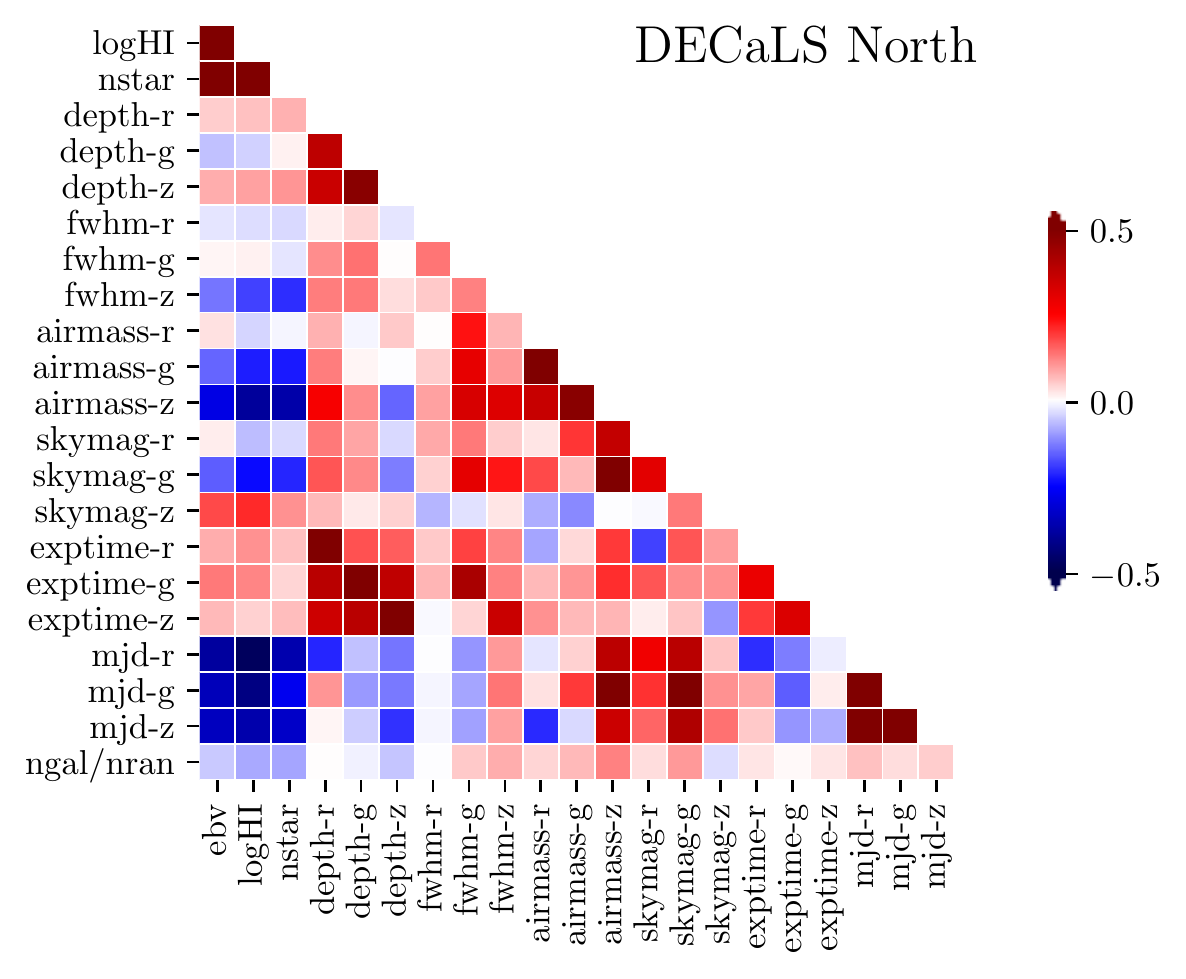}
	\includegraphics[width=\columnwidth]{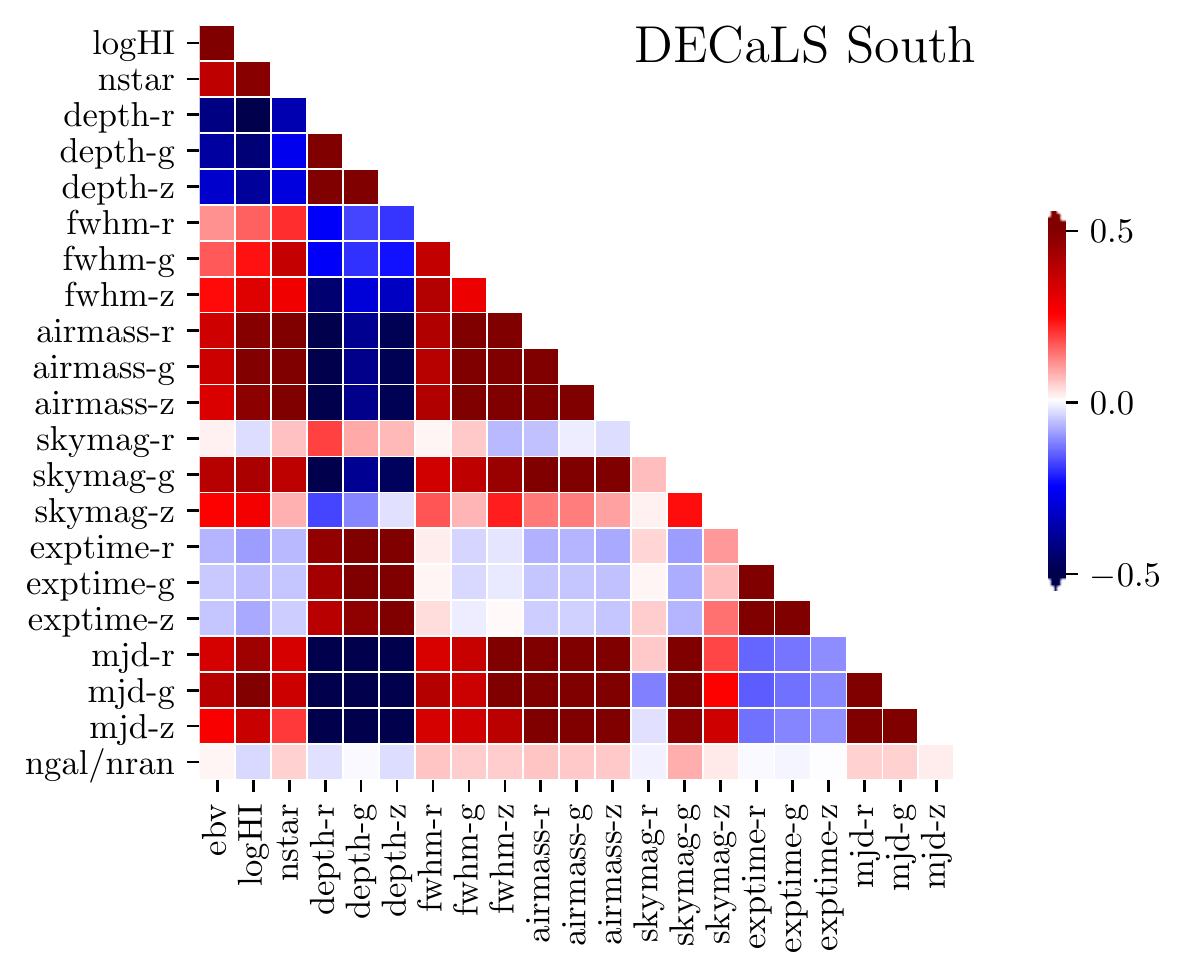}
	\includegraphics[width=\columnwidth]{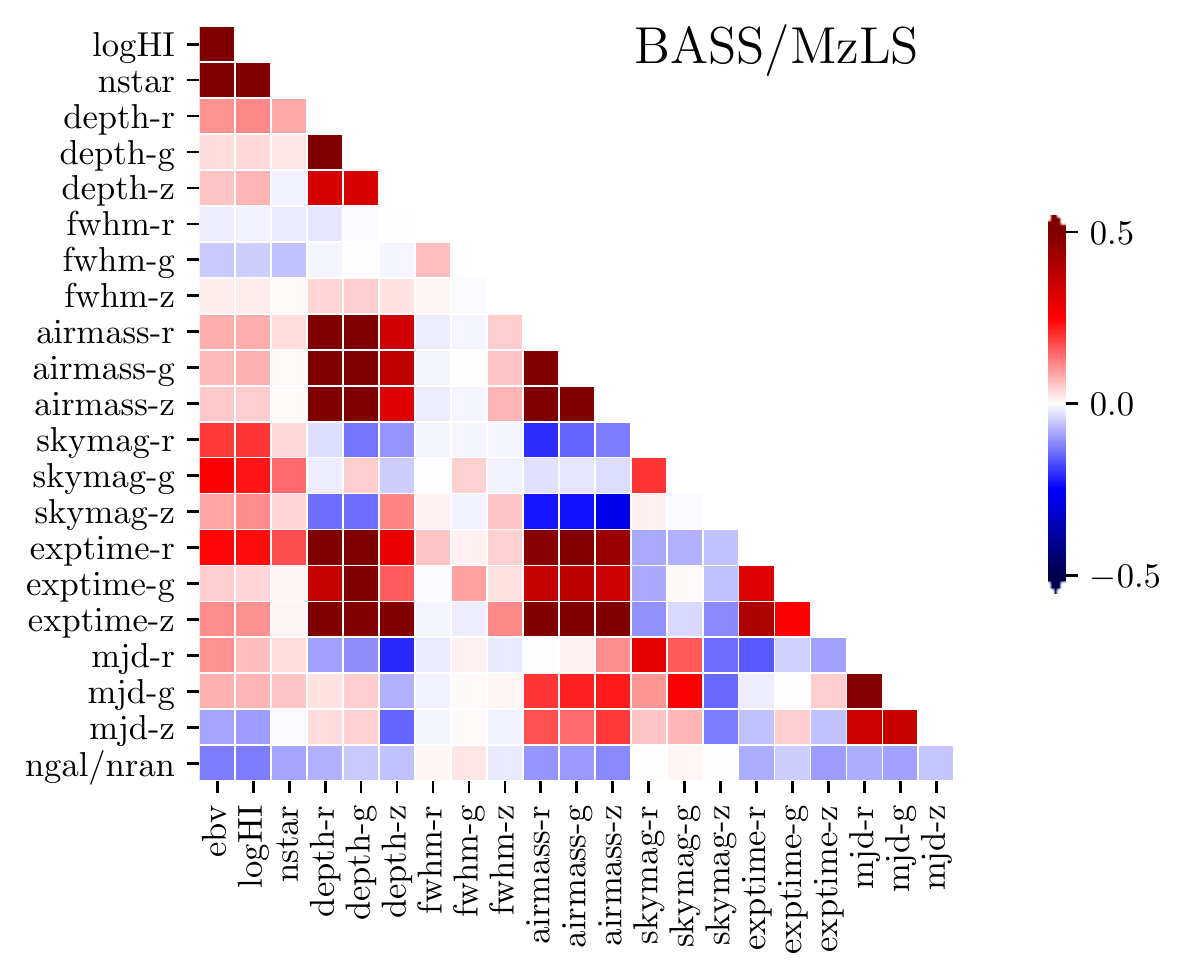}
    \caption{Correlation coefficients between each imaging systematics and the observed galaxy density. Top: DECaLS-North, middle: DECaLS-South and bottom: BASS/MzLS.}
    \label{fig:NN} 
\end{figure}

To estimate the linear correlation between each pair of the imaging attributes and the galaxy density, we compute the Pearson correlation coefficient (PCC) defined by:
\begin{equation}
r_{x,y} = \frac{C_{x,y}}{\sqrt{C_{xx}C_{yy}}}
\end{equation}
where $C(x,y)$ corresponds to the covariance between $x$ and $y$ across all pixels. Fig.~\ref{fig:NN} shows the colour-coded Pearson correlation matrix between each pair of the imaging attributes and the galaxy density (ngal/nran) where the top panel corresponds to DECaLS-North, the middle panel to DECaLS-South and the bottom panel to BASS/MzLS. First, we confirm that we cannot neglect the correlations between the imaging quantities and the complex shape of the overall matrix needs to be taken into account when correcting for the variations of the galaxy density with these systematics. We can also see different behaviours across the imaging surveys. For instance, as in~\citet{Rezaie+19} we also find an anti-correlation between the Galactic foregrounds (stellar density, neutral hydrogen column density and Galactic extinction) and the observed galaxy density in the NGC with BASS/MzLS and DECaLS-North but we find a positive correlation in DECaLS-South.
As expected, the observed galaxy density is also anti-correlated with the depth, because we select more low-$z$ objects with shallower imaging, but the amplitude of the correlation in each band can vary between the surveys.

\begin{figure*}
    \centering
    \includegraphics[page=1, width=\textwidth, clip]{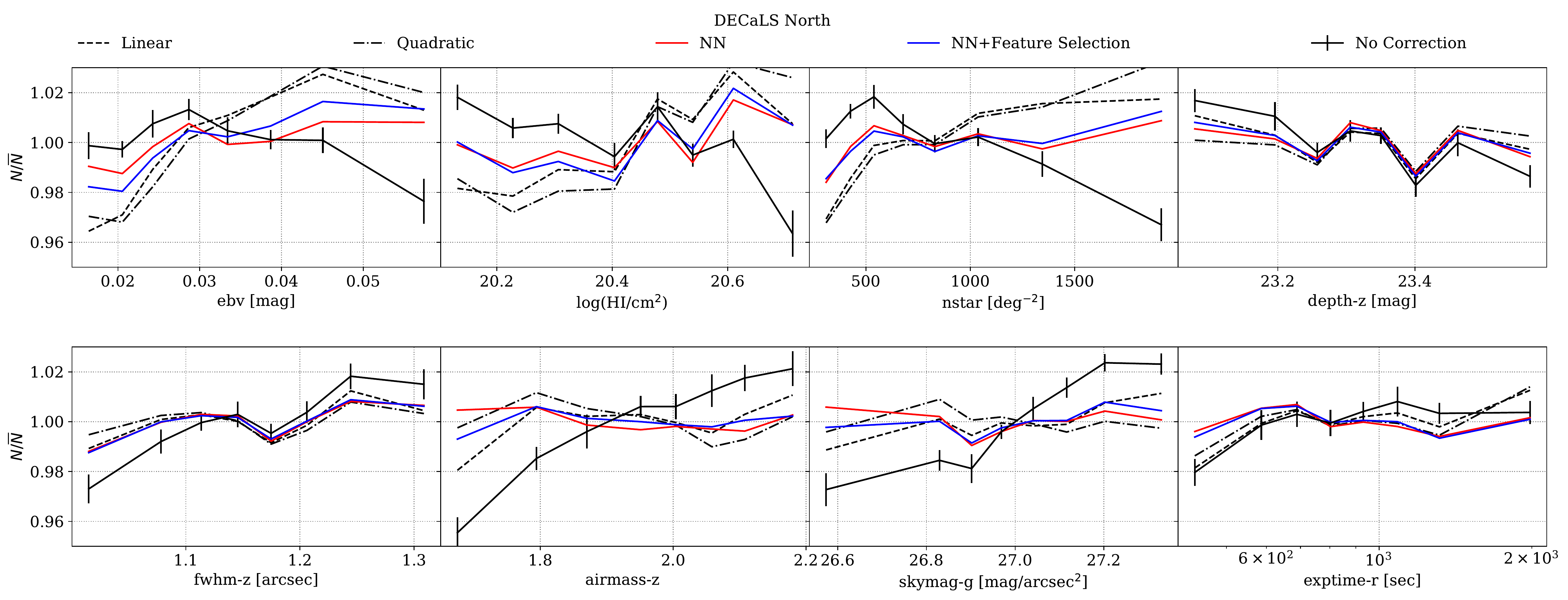}
    \includegraphics[page=2, width=\textwidth, clip]{Figures/Figure8_photo-syst.pdf}
    \includegraphics[page=3, width=\textwidth, clip]{Figures/Figure8_photo-syst.pdf}
    \caption{Target density variation of the photometric sample with imaging systematics before applying weights (black) and after assuming a linear (dotted), quadratic (dashed) or neural network based (red) relation between the observed galaxy density and the potential systematics. Top: DECaLS-North, middle: DECaLS-South and bottom: BASS/MzLS.}
    \label{fig:photo-syst}
\end{figure*}

Because of the different behaviours, the weights are derived by fitting the entire DESI footprint split into BASS/MzLS, DECaLS-North and DECaLS-South. Fig.~\ref{fig:photo-syst} shows the target density variations with the most important imaging systematics where the top panel displays the variation in DECaLS-North, the middle panel in DECaLS-South and the bottom panel in BASS/MzLS. In all cases, the solid black curve corresponds to the galaxy density without any correction; the dotted  curve is using a multi-variate regression technique assuming a linear relation between the observed galaxy density and the systematics; the dashed curve assumes a quadratic relation and the solid red line corresponds to the case after applying the neural network.
As seen with the PCC, for the same systematic quantity we can have different trends of the observed galaxy target density in each region. However, in the three regions, only the NN approach enables correction for the non-linear variations with systematics and recovery of a constant target density. In Section~\ref{sec:clust-cross} we will show the impact of each set of weights on the projected cross-correlation function.
Another step called `feature selection' was also added in~\citet{Rezaie+19} in order to reduce redundancy among the imaging attributes and to avoid over-fitting the cosmological signal. We also apply this step and derive another set of NN weights after applying the feature selection which corresponds to the blue curve in Fig.~\ref{fig:photo-syst}. In our case, it has a marginal impact on the target density variation, the projected cross-correlation function and the BAO constraints. More details can be found in Appendix~\ref{sec:appendixA}.

In what follows, in order to avoid mixing the surveys and because the analysis is limited to the eBOSS footprint, we remove the region at Dec $< 32.275$ deg~\footnote{\url{https://github.com/desihub/desitarget/blob/master/py/desitarget/io.py\#L95}} in both the spectroscopic and photometric samples to consider BASS-MzLS only in the NGC. We also remove the DES region in the SGC (Dec $< 5$ deg) as it is deeper than DECaLS. Given that we restrict to the eBOSS footprint, the removed regions have small areas so the statistical precision is only slightly affected but in Appendix~\ref{sec:appendixB}, we show that mixing these surveys degrades the cosmological signal significantly.


\section{Methodology}
\label{sec:method}

\subsection{Clustering estimators}
\label{sec:estimator}
Given that random catalogues are made available for both spectroscopic and photometric objects, we can use the Landy-Szalay estimator~\citep{LandySzalay93}:
\begin{equation}
     w_{\theta}(\mathbf{R})=\frac{D_{1}D_{2}(\mathbf{R})-D_{1}R_{2}(\mathbf{R})- D_{2}R_{1}(\mathbf{R})+R_{1}R_{2}(\mathbf{R})} {R_{1}R_{2}(\mathbf{R})}
\end{equation}
where $DD$, $DR$ and $RR$ are the paircounts between data-data, data-random and random-random respectively at average separation $R$.
For the spectroscopic dataset, we generate a catalogue of randoms 25 times larger than the eBOSS quasar catalogue following the same methodology as in~\citet{Ross+20} for the official eBOSS DR16 quasar analysis.
For the photometric dataset, we take the randoms made for the Legacy Surveys~\footnote{\url{http://www.legacysurvey.org/dr8/files/random-catalogs}} and given the already high density sampling, we generate a catalogue of randoms only 5 times larger than the target catalogue.

We modify the publicly available code TWOPCF\footnote{\url{https://github.com/lstothert/two\_pcf}} to include the calculation of the cross-correlation function binned in transverse comoving separation. 

\subsection{Covariance matrix}
\label{sec:covmatrix}

\begin{figure}
	\centering
    \includegraphics[width=\columnwidth]{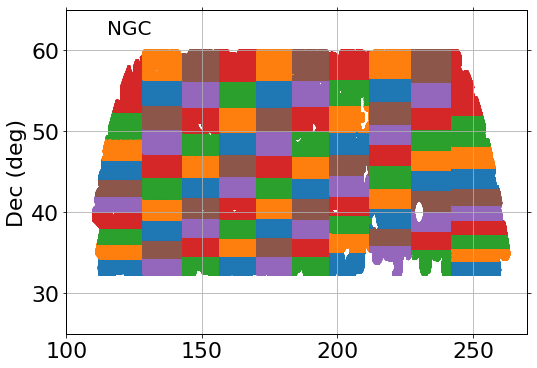}
    \includegraphics[width=\columnwidth]{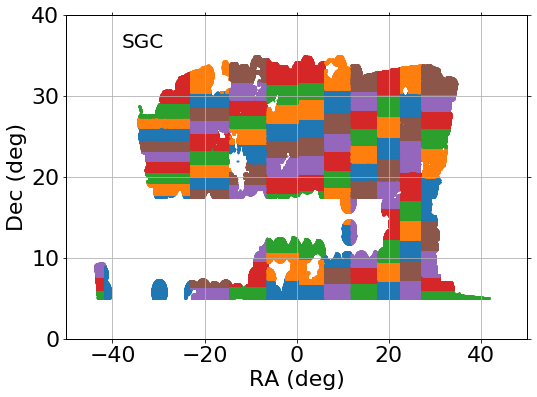}
    \vskip -0.2cm
    \caption{Footprint of the NGC (top panel) and SGC (bottom panel) where colours indicate the 100 jackknife regions.}    
    \label{fig:jackknife} 
\end{figure}

The code can also calculate jackknife errors in a single loop over the galaxy pairs which makes this calculation very efficient. Therefore, we use the jackknife method to estimate our covariance matrix~\citep[for a review on the error estimation methods, see][]{Norberg+09}.
The covariance matrix is given by:
\begin{equation}
C_{i,j} = \frac{N_{j}-1}{N_{j}} \sum_{n=1}^{N_{j}} [\xi_{l,n}(s_{i}) - \bar{\xi_{l}}(s_{i})] \, [\xi_{l',n}(s_{j}) - \bar{\xi_{l'}}(s_{j})]
\end{equation}
where $N_{j}$ is the number of jackknife realizations.
We divide the footprint into 100 independent sub-regions for both the spectroscopic and photometric sample as showed in Fig.~\ref{fig:jackknife}.
To do so, we create regions of similar area by splitting the survey with straight line cuts in RA and then Dec such that each region contains the same number of points in the random catalogue.

\begin{figure}
	\centering
    \includegraphics[width=\columnwidth]{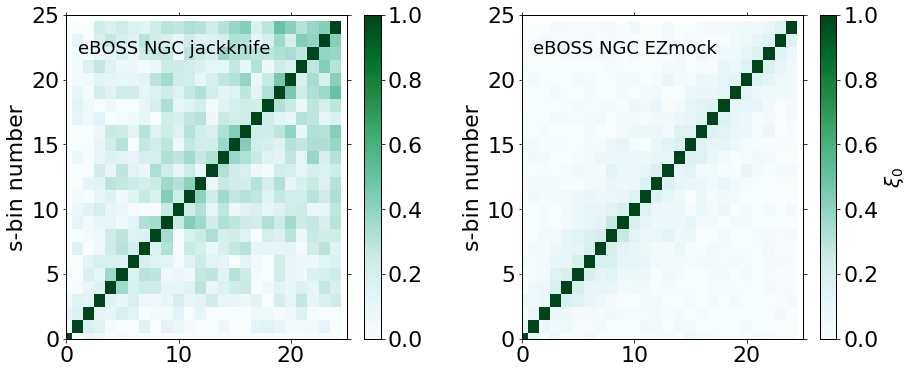}
    \includegraphics[width=\columnwidth]{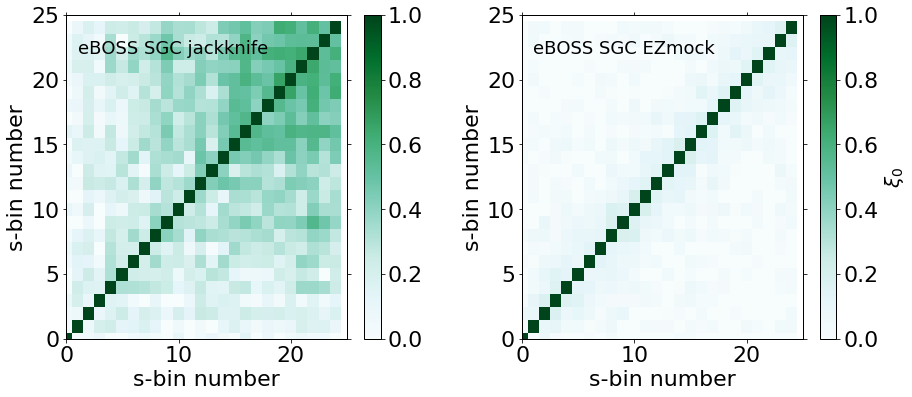}
    \caption{Correlation matrix obtained from the 100 jackknife regions and used to fit the monopole of the eBOSS quasars auto-correlation function in 25 bins of width 8 h$^{-1}$Mpc between 0 and 200 h$^{-1}$Mpc and in the redshift range $0.8 \leq z \leq 1.5$.}
    \label{fig:auto-corr-matrix} 
\end{figure}

\begin{figure}
	\centering
    \includegraphics[scale=0.4]{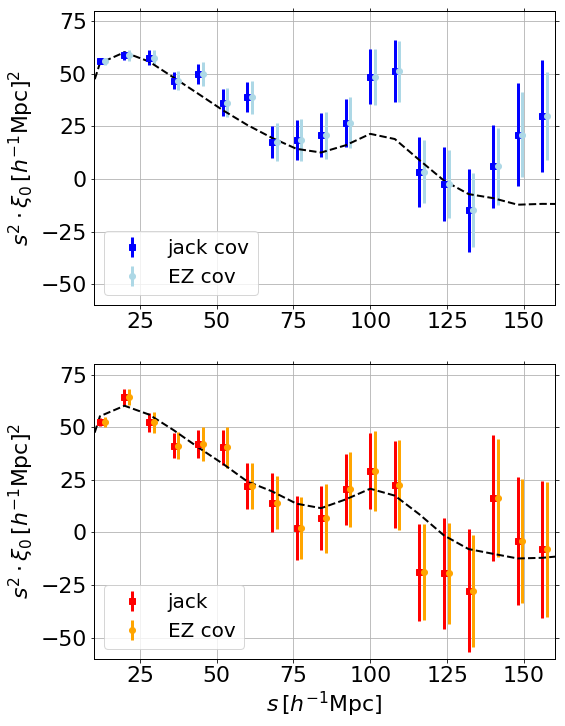}
    \caption{Monopole of the auto-correlation of eBOSS DR16 quasars in the NGC (top) and SGC (bottom) with error bars coming from the 100 jackknife realisations and from the 1000 EZ mocks.}
    \label{fig:auto-jack-EZ} 
\end{figure}

We then compute the corresponding covariance matrix for the monopole of the eBOSS quasars auto-correlation function and the projected cross-correlation function \com{separately} after applying their respective weights.
For the auto-correlation function, in the redshift range $0.8 \leq z \leq 1.5$, we can compare the covariance matrix and diagonal elements obtained from the 100 jackknife regions and the 1000 eBOSS QSO EZ mocks~\citep{Zhao+20} which are used in the cosmological analysis of the eBOSS DR16 quasars~\citep{Hou+20,Neveux+20}. These mocks are based on the Effective Zel'dovich approximation following the method developed in~\citet{Chuang+15}; they use 7 simulation snapshots to create a lightcone and they are tuned to match the clustering of the final DR16 quasar catalogues. 
Fig.~\ref{fig:auto-corr-matrix} displays the correlation matrix for the monopole of the auto-correlation function of eBOSS quasars in $0.8 \leq z \leq 1.5$ obtained from the 100 jackknife realisations (left panel) and the 1000 EZ mocks (right panel). The top (respectively bottom) row shows the results for the NGC (respectively SGC). As expected, the correlation matrix from the jackknife method is noisier as it is limited by the number of independent realisations we can create from the survey. We can also compare the diagonal elements as shown in Fig.~\ref{fig:auto-jack-EZ} where we can see the monopole of the auto-correlation function for the NGC (top) and the SGC (bottom) obtained from both methods. They give very similar errors on the measurement while they rely on different assumptions, which gives confidence that both techniques provide a reasonable estimate of the error bar. \com{We also checked that the covariance matrix was stable when varying the number or locations of the jackknife regions.}

\begin{figure}
	\centering
    \includegraphics[width=0.8\columnwidth]{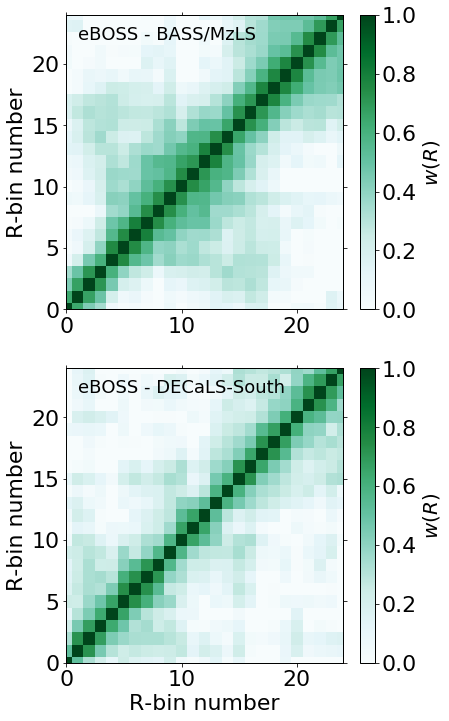}
    \vskip -0.2cm
    \caption{Correlation matrix obtained from the 100 jackknife regions and used to fit the projected cross-correlation in 25 bins of width 8 h$^{-1}$Mpc between 0 and 200 h$^{-1}$Mpc.}
    \label{fig:cross-corr-matrix} 
\end{figure}

The correlation matrix of the projected cross-correlation function in the NGC / BASS-MzLS (top panel) and the SGC / DECaLs (bottom panel) is shown in Fig.~\ref{fig:cross-corr-matrix}. In this case, we do not have available mocks with the properties of both the photometric and spectroscopic samples so we can only use the jackknife method to fit the data. We checked that the correlation matrix was robust with respect to binning and systematics weights for the photometric sample and in Section~\ref{sec:bao-results} we will also show the BAO results when fitting each individual jackknife region and taking the mean.

\subsection{BAO fitting procedure}
\label{sec:baofit}

We use a similar BAO fitting procedure for the auto-correlation function of the eBOSS DR16 quasars and the projected cross-correlation function, except for the template.
We follow the same methodology as for the BAO analysis in the DR14 eBOSS quasars~\citep{DR14-bao} and in the BOSS LRG sample~\citep[DR10/DR11, ][]{Anderson+14} and ~\citep[DR9, ][]{Anderson+12}.
\begin{enumerate}
\item compute the two-point statistics (using the LS estimator for the correlation function as described in Section~\ref{sec:estimator})
\item generate a template BAO feature ($\xi_{\rm temp}$ or $w_{\rm temp}$) using the linear power spectrum $P_{\rm lin}(k)$ obtained from CAMB assuming a fiducial cosmological model
\item generate a template without the BAO feature where $P_{\rm nw}(k)$ (`nw' stands for `no wiggle' which corresponds to no BAO feature in $k$-space) obtained from the fitting formulae in~\cite{EisensteinHu98} using the same fiducial cosmological model:
\end{enumerate} 

We can then model the auto-correlation function following~\cite{Xu+12}:
\begin{equation}
\xi^{\rm mod}(s) = B_{0}\, \xi_{\rm temp}(\alpha, s) + A_{1} + A_{2} / s + A_{3} / s^{2}
\end{equation}
where $B_{0}$ is a multiplicative constant allowing for an unknown large-scale bias and $A_{1,2,3}$ are the coefficients of the additive polynomial function to make the results insensitive to shifts in the broad-band shape of the measured correlation function. \com{As in \citet{DR14-bao}, we apply a Gaussian prior of width 0.4 around the  $B_{0}$ value found when fitting the template to the data in $10 < s,R < 80$
~h$^{-1}$Mpc without including the broad-band terms.}

The BAO template for the correlation function is obtained by Fourier transformation of the power spectrum:
\begin{equation}
\xi_{\rm temp}(s) = \int \frac{k^{2}dk}{2\pi^{2}} P_{\rm temp}(k) j_{0}(ks) e^{-k^{2}a^{2}}
\end{equation}
where the exponential term has been introduced to damp oscillatory patterns associated with the Bessel function $j_{0}$ at high-$k$ and induce better numerical convergence~\citep{Anderson+14}. The exact damping scale is not important, in this analysis it is set to $a=1 h^{-1}{\rm Mpc}$. 

The template for the power spectrum is given by:
\begin{equation}
P_{\rm temp}(k) = P_{\rm nw}(k) \left[ 1 + \left( \frac{P_{\rm lin}(k)}{P_{\rm nw}(k)} - 1 \right) e^{-\frac{1}{2}k^{2}\Sigma^{2}_{\rm nl}} \right]
\end{equation}
where the BAO signature in linear theory is described by the oscillatory pattern in the $O_{\rm lin}(k) = P_{\rm lin}(k) / P_{\rm nw}(k)$ and the $\Sigma^{2}_{\rm nl}$ term is used to damp the acoustic oscillations in the linear theory power spectrum to account for the effects of non-linear evolution of the density field. As in~\cite{DR14-bao}, we use $\Sigma^{2}_{\rm nl} = 6 \, [h^{-1}{\rm Mpc}]^{2}$ but previous studies showed that the results are insensitive to this choice, as also confirmed in galaxy samples.

For the cross-correlation function template, we use equation~\ref{eq:binned-wR} with both the linear $P_{\rm lin}(k)$ \com{and $P_{\rm temp}(k)$ with the damping term to account for non-linear effects.}

For both clustering statistics, we determine how different the BAO scale is in the clustering measurements compared to its location in a template generated using our fiducial cosmology. The observed BAO position can differ from the one in the template because of two main effects. The first effect is related to the fact we do not know the BAO position in the true intrinsic primordial power spectrum, therefore to account for this in the template we include a multiplicative shift which depends on the ratio $r_{\rm drag}/r_{\rm drag}^{\rm fid}$, where $r_{\rm drag}$ is the sound horizon at the drag epoch and corresponds to the expected location of the BAO feature in comoving distance units. The second effect is due to the fact we need to assume a fiducial cosmological model to convert angles and redshift from the catalogue into comoving coordinates. If the true cosmology is different than the one we assumed, the inferred clustering will contain detectable distortions, in addition to the redshift space distortions due to peculiar velocities. This effect is known as the Alcock-Paczynski effect~\citep{AP}.
By introducing two shift parameters, $\alpha_{\parallel}$ and $\alpha_{\perp}$, we can account for this dilation of scales in the direction along and perpendicular to the line-of-sight The parameters $\alpha_{\parallel}$ and $\alpha_{\perp}$ can be related to the expansion rate $H(z)$ and the comoving angular diameter distance $D_{\rm M}$ through:
\begin{equation}
\alpha_{\parallel}=\frac{H^{\rm fid}(z)r_{\rm drag}^{\rm fid}}{H(z)r_{\rm drag}},\qquad \alpha_{\perp}=\frac{D_{\rm M}(z)r_{\rm drag}^{\rm fid}}{D_{\rm M}^{\rm fid}(z)r_{\rm drag}}
\label{eq:H-DA}
\end{equation}
However, given the low statistical precision of the quasar sample in the redshift ranges we consider, it is more optimal to fit an isotropic shift $\alpha_{\rm iso}$ and constrain the spherically-averaged distance $D_{\rm V}$ as in~\citet{DR14-bao}:
\begin{equation}
D_{\rm V}=\left[ (1+z)^2cz\frac{D_{\rm M}^2}{H} \right]^\frac{1}{3}
\label{eq:DV}
\end{equation} 
where $c$ is the speed of light and  $\alpha_{\rm iso}$ is thus defined by:
\begin{equation}
\alpha_{\rm iso}=\frac{D_{\rm V}(z)r_{\rm drag}^{\rm fid}}{D_{\rm V}^{\rm fid}(z)r_{\rm drag}}\,.
\end{equation}

Therefore, the auto-correlation function measurements will enable the constraint of $D_{\rm V}(z)$ by fitting an isotropic shift $\alpha_{\rm iso}$ while the projected cross-correlation function will put constraints on the comoving angular diameter distance $D_{\rm M}(z)$ by fitting a transverse shift $\alpha_{\rm \perp}$.


\subsection{Effective redshift}
\label{sec:effective-z}

\com{It is common in standard clustering analyses to approximate the effects of fiducial cosmology as a single rescaling of the cosmological parameters. The redshift range that the eBOSS quasars span is broad, between $z=0.8$ and $z=2.2$ and therefore the redshift evolution is more important. Recent techniques using a redshift-weighting have been developed to account for the redshift evolution of the parameters. In
~\citet{Zhu+18}, they performed a BAO analysis of the eBOSS DR14 QSO sample and the constraint they obtained on the spherically-averaged distance $D_{\rm v}$ with and without redshift weighting differs by less than 1\%. So we can neglect this effect for this analysis where the redshift range is also smaller and consider a single effective redshift.} 
In order to match the definition used for the clustering analysis of the eBOSS DR16 quasars in $0.8 \leq z \leq 2.2$~\citep{Hou+20,Neveux+20}, we define the effective redshift by:
\begin{equation}
z_{\rm eff} = \frac{\sum_{i,j} w_{\mathrm{tot},i} w_{\mathrm{tot},j} (z_{g,i} + z_{g,j})/2}{\sum_{i,j} w_{\mathrm{tot},i} w_{\mathrm{tot},j}}
\label{eq:zeff-spectro}
\end{equation}
where the sum is performed over all galaxy pairs between 0~$h^{-1}{\rm Mpc}$ and 200~$h^{-1}{\rm Mpc}$.
We use this definition for the auto-correlation function.

For the cross-correlation function, we do not know the redshift distribution for the photometric sample. Nevertheless, we can make different assumptions and study their impact on the definition of the effective redshift and thus on the cross-correlation function.
If we assume the redshift distribution of the photometric sample is flat, then the effective redshift depends only on the distribution of the spectroscopic sample such that $z_{\rm eff,1}$ is given by equation~\ref{eq:zeff-spectro}.
We also consider another definition, $z_{\rm eff,2}$ where we account for the shape of the redshift distribution of the photometric galaxies by weighting the spectroscopic redshift distribution with the signal-to-noise ratio of the cross-correlation signal. \com{To do so, we divide the quasar sample into redshit bins of width $\Delta z=0.1$.}
Finally, we also compute the effective redshift by using the HSC-PDR2 photometric redshifts~\citep{HSC-pdr2-photoz} when available with bins of width $\Delta z=0.1$.
For the two redshift ranges we consider, we obtain the following:
\begin{eqnarray}
0.6 \leq z \leq 1.2: z_{\rm eff,1} = 0.95 \; , \; z_{\rm eff,2}=0.92 \; , \; z_{\rm eff,3}=0.93 \nonumber \\
0.8 \leq z \leq 1.5: z_{\rm eff,1} = 1.19 \; , \; z_{\rm eff,2} = 1.10 \; , \; z_{\rm eff,3} = 1.19 \nonumber
\end{eqnarray}
\com{The three definitions are in good agreement with relatively close values. Moreover, because we measure a distance relative to the fiducial assumption that has been measured and not an absolute distance, there is indeed some systematic uncertainty on how the impact of the effective redshift choice plays out when testing cosmology, but it is minor.}. 
Therefore, in Section~\ref{sec:bao-cross}, we measure the projected BAO scale assuming $z_{\rm eff,1}$ only but we checked that using $z_{\rm eff,2}$ leads to the same results.


\subsection{Parameter inference}
\label{sec:parameter}

We extract the results of the fitting of either the monopole of the auto-correlation function or the projected cross-correlation function by minimising the $\chi^{2}$ defined by:
\begin{equation}
\chi^{2} = (\xi^{\rm Data} - \xi^{\rm Model}) C^{-1} (\xi^{\rm Data} - \xi^{\rm Model})^{T}
\end{equation}
where $\xi^{\rm Data}$ corresponds to the measurement, $\xi^{\rm Model}$ to the associated theoretical prediction, and $C^{-1}$ the inverse of the estimated covariance matrix. 
When using the 1000 eBOSS QSO EZ mocks~\citep{Zhao+20} to obtain a covariance matrix in $0.8 \leq z \leq 1.5$, we include the Hartlap correction~\citep{Hartlap+07} due to finite number of mocks and number of bins in the analysis that can bias the measurements:
\begin{equation}
C^{-1}_{\rm unbiased} = (1 - D) C^{-1}_{\rm mock} \, \, {\rm with} \, \, D = \frac{N_{b}+1}{N_{m}-1}
\end{equation}
where $N_{b}$ is the total number of bins in the measurements and $N_{m}=1000$ is the number of realizations.

We use the public code BAOfit\footnote{\url{https://github.com/ashleyjross/BAOfit}} to perform the BAO fitting for both the auto and cross-correlation functions (using our template for the fit to the cross-correlation).

\section{Results}
\label{sec:results}

\subsection{Clustering measurements}

\subsubsection{Auto-correlation function}
\label{sec:clust-auto}
\begin{figure}
	\centering
    \includegraphics[width=\columnwidth]{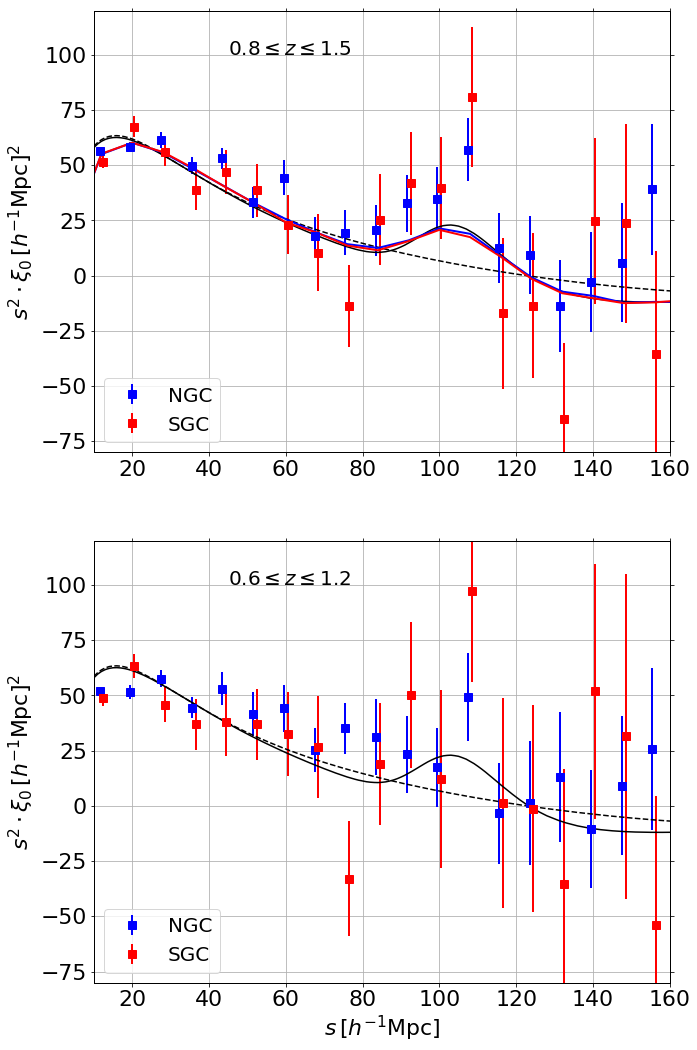}
    \vskip -0.2cm
    \caption{Monopole of the auto-correlation function of the eBOSS DR16 quasars for the NGC (red squares) and SGC (blue squares) for quasars in $0.8 \leq z \leq 1.5$ (top) and $0.6 \leq z \leq 1.2$ (bottom). The solid curves display the mean of the 1000 EZ mocks. The data in each region are consistent with each other and with the mean of the mocks.}
    \label{fig:clustering-auto} 
\end{figure}

We measure the monopole of the auto-correlation function of eBOSS DR16 quasars in the NGC-BASS/MzLS and SGC-DECaLS regions separately and for the two redshift ranges we consider. To do so, we follow the methodology in~\citet{DR14-bao} where we calculate the auto-correlation function $\xi(s,\mu)$ in evenly spaced bins in $s$ from 0 to 200~$h^{-1}$Mpc with a bin width of 8~$h^{-1}$Mpc and 0.01 in $\mu$. The multipoles of the auto-correlation function are then determined by:
\begin{equation}
    \xi_{l}(s) = \frac{2l+1}{2} \sum_{i=1}^{100} 0.01\, \xi(s,\mu_i)L_{l}(\mu_i)
\end{equation}
where $\mu_i = 0.01i - 0.005$ and $L_{l}$ is the Legendre polynomial of order $l$. In this work, we use only the monopole ($l=0$). This definition of the monopole ensures an equal weighting as a function of $\mu$ which thus corresponds to a truly spherically averaged observable.
Fig.~\ref{fig:clustering-auto} displays the spherically-averaged redshift-space auto-correlation function in the NGC (red squares) and SGC (blue squares) for quasars in $0.8 \leq z \leq 1.5$ (top) and $0.6 \leq z \leq 1.2$ (bottom). The solid curves in the top panel show the mean of the 1000 EZ mocks available in this redshift range for the NGC (blue) and SGC (red). The data in each region is consistent with each other, and with the mean of the mocks for the redshift range $0.8 \leq z \leq 1.5$ where EZ mocks are available.

\subsubsection{Projected cross-correlation function}
\label{sec:clust-cross}

As for the auto-correlation function, we compute the projected cross-correlation function in evenly spaced bins in $s$ from 0 to 200~$h^{-1}$Mpc with a bin width of 8~$h^{-1}$Mpc.
We first investigate the impact of the imaging weights used for the photometric sample on the projected cross-correlation function. As described in Section~\ref{sec:photo-data}, we derived the three sets of weights separately for the two regions, BASS/MzLS and SGC-DECaLS, which together comprise our full DESI footprint. Each set is based on a different assumption to model the dependence between the observed target density and the imaging attributes: linear relation, quadratic relation, neural network.

\begin{figure}
	\centering
    \includegraphics[width=\columnwidth]{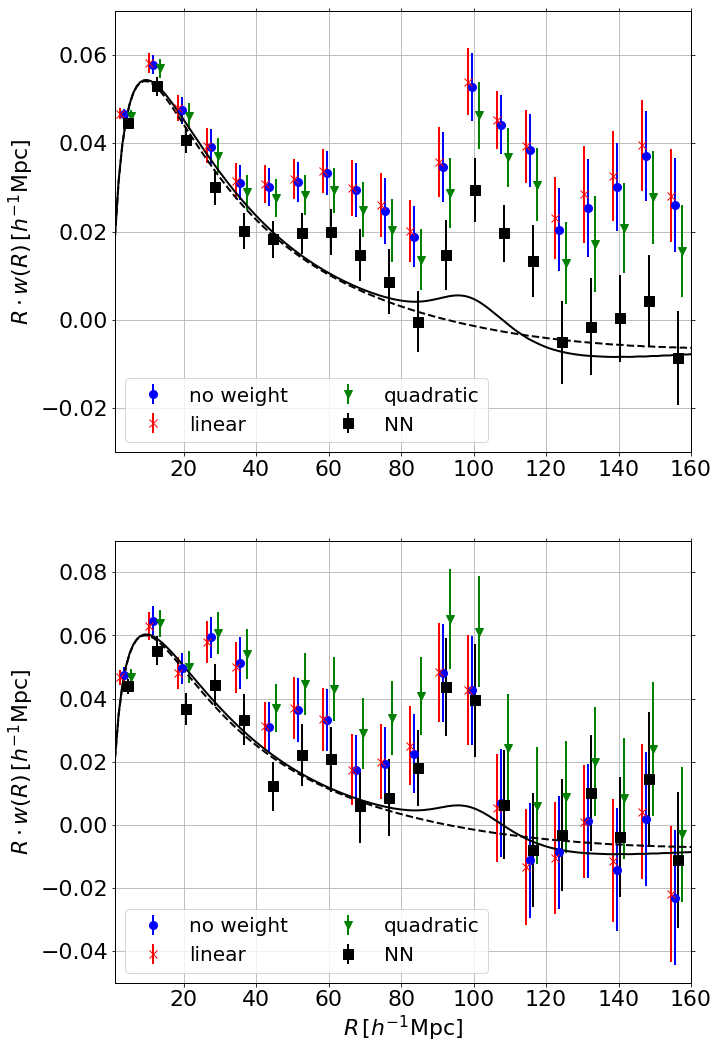}
    \vskip -0.2cm
    \caption{$w(R)$ for NGC-BASS/MzLS (top) and SGC-DECaLS (bottom) regions with and without weights assuming different relations between the observed galaxy density and the potential systematics. \com{The solid (resp. dashed) curve shows the template for the projected corrrelation function with (resp. without) the BAO feature.} For simplicity, we show the redshift range $0.8 \leq z \leq 1.5$ only but we find a similar behaviour for $0.6 \leq z \leq 1.2$.}
    \label{fig:clustering-cross-weight} 
\end{figure}

The projected cross-correlation function in the NGC-BASS/MzLS (top) and SGC-DECaLS (bottom) regions for each type of weights is shown in Fig.~\ref{fig:clustering-cross-weight} for the cross-correlation with quasars in $0.8 \leq z \leq 1.5$. The blue dots correspond to the case without applying imaging weights to the photometric sample, the red dots shows the case after applying linear weights, the green dots after applying quadratic weights and the black dots correspond to the case after applying neural network weights.
In general, applying the imaging weights improves the agreement with the model (solid black curve with BAO and dashed black curve without) and as expected the weights based on the neural network provide the best improvement.
\com{However, the agreement is less good around $50 \, h^{-1}$Mpc (especially in the NGC) and on scales above $\sim 110 \, h^{-1}$Mpc in the NGC-BASS/MzLS region suggesting that there is some remaining systematics the neural network did not capture and/or which is not contained in the imaging attributes used to train the neural network.} We found a similar behaviour both for the impact of the different weights and the potential remaining systematics in the NGC when cross-correlating with quasars in $0.6 \leq z \leq 1.2$. \com{We checked that the correlation coefficient between these points is high, typically about 0.75, meaning that the same trend between the points is expected. We also computed the projected correlation function of the quasar sample and the angular correlation function of the photometric sample and nothing unusual was found.}

\begin{figure}
	\centering
    \includegraphics[width=\columnwidth]{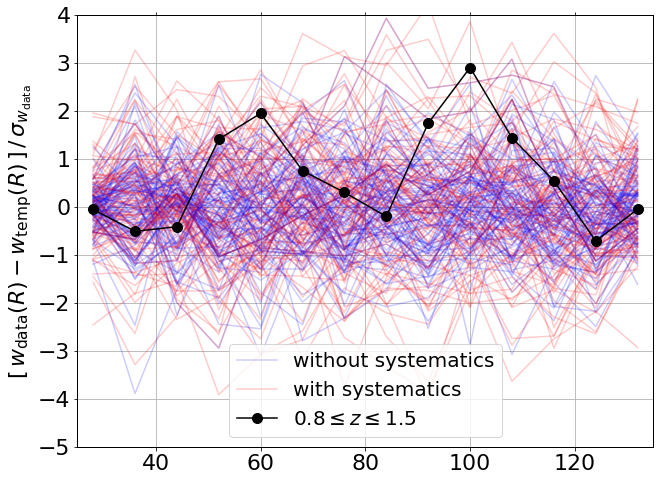}
    \vskip -0.2cm
    \caption{Difference in standard deviation between the projected cross correlation function of the data and the template given by equation~\ref{eq:binned-wR}. The black dots correspond to the weighted mean of the data, the blue curves show the 100 DESI EZ mocks without imaging systematics and the red ones with systematics.}
    \label{fig:data-vs-temp} 
\end{figure}

We also note that the amplitude of the BAO feature is more pronounced in the data than in the template. In order to investigate the validity of the template for the projected cross-correlation function given by equation~\ref{eq:binned-wR}, we use the DESI ELG EZ mocks based on the Effective Zel'dovich approximation following the method developed in~\citet{Chuang+15}. We need to have both the spectroscopic and photometric samples in the same mock realisation. Therefore we use the DESI ELG EZ mocks which have been tuned to match the 
target density of the main DESI selection which is about 2400~deg$^{-2}$ over the entire DESI footprint and use only the angular coordinates for the photometric sample. In order to mimic the spectroscopic quasar sample, in each mock we consider objects in $0.8 \leq z \leq 1.5$ and randomly downsample them to reach the target density of the eBOSS quasars in this redshift range. By doing so, we do not expect the bias of the samples to be exactly the same as the one of the data, especially for the spectroscopic sample as we know quasars are more biased tracers than ELG. A better way of creating a higher bias sample like the quasar one would be to select the objects based on the background density but the density information is not available in this version of the DESI EZ mocks.
\com{We also imprint the inhomogeneities observed in the the number density of the ELGs selected from the DESI Legacy Imaging Surveys. The depth of the survey varies across the footprint which introduces a fluctuation in the number of ELGs detected across the sky. This artificial fluctuation is also function of redshift. In order to assess the fluctuation imprinted in the redshift distribution of ELGs we use a Monte Carlo approach\footnote{See this notebook for technical details \url{https://github.com/desihub/LSS/blob/master/Sandbox/MCeff.ipynb}}. We take a 3 $\deg^2$ region of DECaLS(DR7) where the photometry is $\sim$1.5 magnitude deeper than the average photometry and which is inside the HSC(DR2) footprint \citep{HSC-pdr2}.
We match that sample with the HSC(DR2), and thus have a sample that we consider as our ``truth" sample, where we have deep $grz$-photometry, along with their errors, and a precise redshift estimation from HSC(DR2).
We obtain our ``truth" redshift distribution by applying the DESI ELG target selection to that sample.
We then imprint the fluctuation in the DESI ELG mocks following these steps:
\begin{itemize}
    \item We divide the DESI footprint in HEALpix pixels and repeat the steps below for each pixel.
    \item We measure the imaging depth in the three photometric bands $g,r$ and $z$.
    \item We take our ``truth" catalogue and add noise to the photometry according to the depth ratio between the considered HEALpix pixel and the original truth one, and we also account for the Galactic extinction.
    \item We then apply the DESI ELG target selection on that degraded photometry, and take the ratio of the redshift distribution obtained from that selection to the ``truth" redshift distribution.
    \item We finally take the DESI ELG mock and select all galaxies in the given HEALpix pixel. We then randomly sub-sample the mock galaxies as a function of redshift using the ratio estimated in the previous step.
\end{itemize}
}
\com{Fig.~\ref{fig:data-vs-temp} shows the difference in standard deviation, $\sigma$, between the projected cross-correlation function of the data, $w_{\rm data}$, and the best-fit model, $w_{\rm temp}$ given by equation~\ref{eq:binned-wR}, for the redshift range $0.8 \leq z \leq 1.5$. The black dots show the result for the weighted mean of the data and each blue (resp. red) curve shows one DESI EZ mock without (resp. with) imaging systematics. We can see the 3$\sigma$ discrepancy at the BAO scale in the data which is counterbalanced by the negative difference in $\chi^{2}$ from correlations between the offset data point, with other data points. However, the behaviour in the data is not completely unusual when compared to the EZ mocks, suggesting that it could be a statistical fluctuation. We also checked the angular clustering of the photometric galaxies and the standard projected auto-correlation function of the quasars and found nothing unusual either.}
\com{In Section~\ref{sec:bao-cross}, we will also present the BAO fits we perform on mocks in order to validate the fitting procedure. When deeper photometric data is available, it would be interesting to look again at this observable. Moreover, when the first set of DESI ELGs spectra are obtained to create a clustering catalogue, it would be worth checking whether the discrepancy at the BAO scale remains.}

\subsection{BAO measurements}
\label{sec:bao-results}

\subsubsection{Auto-correlation function}
\label{sec:bao-auto}

\begin{figure}
	\centering
    \includegraphics[width=\columnwidth]{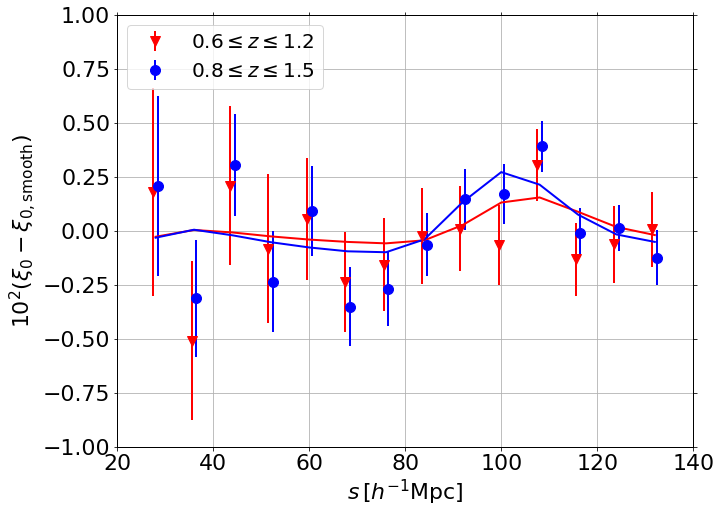}
    \vskip -0.2cm
    \caption{The spherically-averaged BAO signal of the eBOSS DR16 quasars in $0.6 \leq z \leq 1.2$ (red) and $0.8 \leq z \leq 1.5$ (blue). The smooth component of the best-fitting model has been substracted to both the best-fitting model and the measurements in order to isolate the BAO feature.}
    \label{fig:bao-auto} 
\end{figure}

Fig.~\ref{fig:bao-auto} displays the measurement of the BAO feature in the eBOSS DR16 quasar sample in both redshift ranges: $0.6 \leq z \leq 1.2$ (red) and $0.8 \leq z \leq 1.5$ (blue). In each case, we isolate the BAO feature by subtracting the smooth component of the best-fitting model.

Table~\ref{tab:BAO-auto} summarises the results of the BAO fits obtained from the auto-correlation of the eBOSS DR16 quasars in the two redshift ranges we consider for our fiducial configuration and the consistency tests we perform on the data. The fiducial configuration uses a bin width of 8~$h^{-1}$Mpc, a fitting range $20 < s < 140 \, h^{-1}$Mpc, a covariance matrix from 100 jackknife realisations and a total weight defined by equation~\ref{eqn:weight_tot} which is applied to both the data and random catalogues. We note that as shown in the BAO analysis of the eBOSS DR14 quasar sample for instance
~\citet{DR14-bao}, at the precision we are working we do not lose constraining power with bins of this width. We note that the difference in the quoted error between the two redshift ranges is large, which could suggest that the BAO fitting response is sensitive to weak BAO peaks due to low statistics. A similar behaviour was found with the eBOSS DR14 quasar sample when splitting the redshift sample into 2 redshift bins, although the difference in the volume probed by each redshift bin was also bigger~\citep{GilMarin+18}. We also fit the individual jackknife realisations and report the mean result for the value and the error bar. As expected, both the full sample and the mean of the jackknife regions are consistent with each other.
Then, we perform several consistency tests by varying the binning, the fitting range and for $0.8 \leq z \leq 1.5$, we also look at the impact of changing the covariance matrix by using the one from the eBOSS QSO mocks. All the results are consistent with each other within 1$\sigma$, demonstrating the robustness of the BAO feature in the spherically averaged auto-correlation function. 

\begin{table}
    \caption{Results for the isotropic BAO fits to the auto-correlation of the DR16 eBOSS quasars. The fiducial configurations uses data with 8~$h^{-1}$Mpc bin size and centres in the range $20 < s < 140h^{-1}$Mpc and a covariance matrix from 100 jackknife realisations.}
    \label{tab:BAO-auto}
    \begin{tabular}{|c|c|c|}
    \hline
    Configuration & $\alpha_{\rm iso}$ & $\chi^{2}$/d.o.f. \\
    \hline
    \multicolumn{3}{c}{$0.8 \leq z \leq 1.5$} \\
    \hline
    Fiducial                    & 1.013 $\pm$ 0.036 & 12.5/9 \\ 
    mean of the jackknifes                   & 1.014 $\pm$ 0.034 & 13.3/9 \\
    $\Delta s$ = 5$h^{-1}$Mpc   & 1.036 $\pm$ 0.035 & 23.4/19 \\
    EZmock cov fiducial         & 1.005 $\pm$ 0.033 & 14.5/9 \\
    EZmock cov 5$h^{-1}$Mpc     & 1.003 $\pm$ 0.034 & 24.8/19 \\  
    $20 < s < 150h^{-1}$Mpc     & 1.031 $\pm$ 0.036 & 14.7/11 \\  
    $10 < s < 140h^{-1}$Mpc     & 1.029 $\pm$ 0.035 & 15.1/11 \\  
    no $w_{\rm sys}$            & 1.017 $\pm$ 0.035 & 12.4/9 \\   
    NGC                         & 1.009 $\pm$ 0.044 & 9.7/9 \\
    SGC                         & 1.044 $\pm$ 0.063 & 11.0/9 \\ 
    \hline
    \multicolumn{3}{c}{$0.6 \leq z \leq 1.2$} \\
    \hline
    Fiducial                    & 1.003 $\pm$ 0.096 & 6.0/9 \\ 
    mean of the jackknifes                   & 0.999 $\pm$ 0.092 & 6.4/9 \\  
    $\Delta s$ = 5$h^{-1}$Mpc   & 1.019 $\pm$ 0.096 & 18.4/19 \\
    $20 < s < 150h^{-1}$Mpc     & 1.037 $\pm$ 0.106 & 7.0/11 \\
    $10 < s < 140h^{-1}$Mpc     & 1.016 $\pm$ 0.090 & 6.2/11 \\
    no $w_{\rm sys}$            & 0.990 $\pm$ 0.090 & 6.2/9 \\   
    NGC                         & 1.008 $\pm$ 0.098 & 5.3/9 \\
    SGC                         & 1.003 $\pm$ 0.101 & 7.6/9 \\   
    \hline
    \end{tabular}
\end{table} 


\subsubsection{Projected cross-correlation function}
\label{sec:bao-cross}

\begin{figure}
	\centering
    \includegraphics[width=\columnwidth]{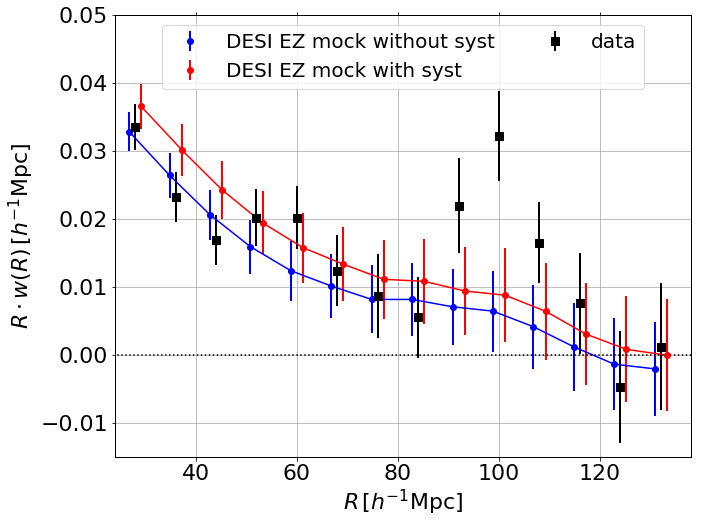}
    \vskip -0.2cm
    \caption{Projected cross-correlation function for the weighted mean of the data (black) compared to the mean of the 100 DESI EZ mocks without (blue) and with (red) imaging systematics. For each EZ mock, the error bar is obtained using 100 jackknife realisations and we show the mean of the error bars for the 100 DESI EZ mocks. For this test, we use quasars in $0.8 \leq z \leq 1.5$.}
    \label{fig:clustering-cross} 
\end{figure}

In order to validate the BAO fitting procedure using the projected cross-correlation function, we first apply the pipeline on the DESI EZ mocks. Fig.~\ref{fig:clustering-cross} displays the projected cross-correlation function of the weighted mean of the data (NGC with BASS/MzLS and SGC with DECaLS-S) in black dots with error bars corresponding to the diagonal elements of the covariance matrix obtained from the 100 jackknife realisations. \com{We also show the mean of the DESI EZ mocks without (blue dots) and with (red dots) systematics while the solid curves show the best-fitting model in each case. In both cases, the error bar corresponds to the mean of the error bars from the 100 DESI EZ mocks. For each EZ mock, the error bar is obtained using 100 jackknife realisations such that the fit to each DESI EZ mock is under the same fitting conditions as for the data.}  Given that the DESI mocks do not completely represent both the photometric and spectroscopic samples by construction, we prefer not to use them to determine the uncertainties on the measurements from the data. We can see that the amplitude of the BAO signal in the mocks is consistent with the template, which confirms the validity of the template. Adding imaging systematics to the mocks yields a constant offset in the projected cross-correlation function but which can be taken into account in the best-fitting model with the normalisation factor. 

\begin{figure}
	\centering
	\includegraphics[width=\columnwidth]{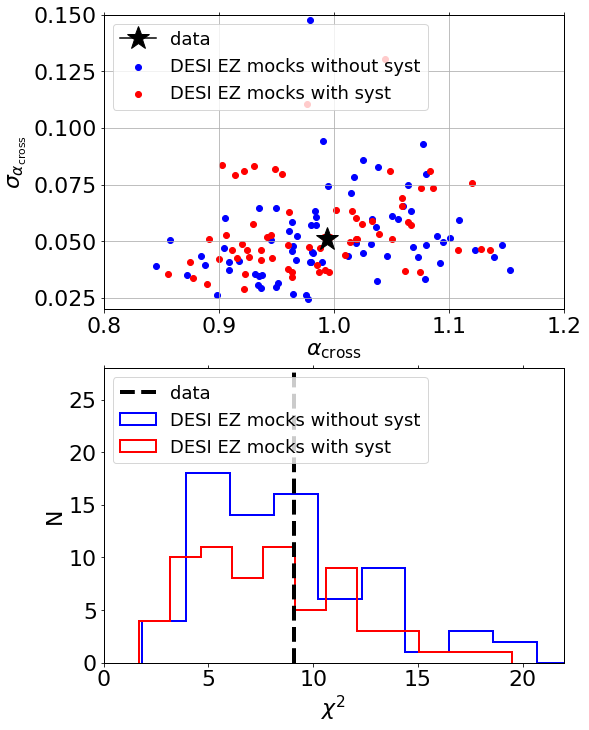}
    \vskip -0.2cm
    \caption{Results of the BAO fitting of the DESI EZ mocks in $0.8 \leq z \leq 1.5$. Top: Value and error bar of $\alpha_{\rm cross}$ for each individual mock without (blue) and with (red) imaging systematics from the Legacy Imaging Surveys. The black star shows the result of the data. Bottom: Distribution of the $\chi^{2}$ for the mocks without (blue) and with (red) imaging systematics compared to the $\chi^{2}$ of the data in dashed black.}
    \label{fig:bao-cross-ez} 
\end{figure}

The results of the BAO fitting are shown in Fig.~\ref{fig:bao-cross-ez} where the top panel displays the value and error obtained on the BAO position for each mock without (blue) and with (red) systematics, compared to the result for the data (black star). 
We fit the individual DESI EZ mocks with their covariance matrix from $20 < s < 140h^{-1}$Mpc. Only mocks with a  `BAO detection' are kept meaning that they have $\Delta \chi^{2} \geq 1$ within $0.8 < \alpha < 1.2$. Over 75\% (resp. 70\%) of the DESI EZ mocks without (resp. with) imaging systematics satisfy this condition. The bottom panel shows the $\chi^{2}$ distribution of the mocks compared to the value for the data. The BAO measurement in the data is consistent with the statistics of the mocks, both in terms of BAO position and $\chi^{2}$, which therefore validates the fitting procedure for the projected cross-correlation function. \com{We also check that the BAO results are robust when using the damping term $\Sigma_{\rm NL}$ which accounts for non-linear effects in the BAO template as in the auto-correlation function. We report the mean value of the 100 individual EZ mocks on the projected BAO parameter $\alpha_{\rm cross}$:
Without systematics
\begin{eqnarray}
 \Sigma^{2}_{\rm NL} &= 0 \; {\rm [h}^{-1} {\rm Mpc]}^{2}: \; \alpha_{\rm cross} = 0.998 \pm 0.051 \nonumber \\
\Sigma^{2}_{\rm NL} &= 6 \; {\rm [h}^{-1} {\rm Mpc]}^{2}: \; \alpha_{\rm cross} = 1.001 \pm 0.058 \nonumber
\end{eqnarray}
With systematics
\begin{eqnarray}
\Sigma^{2}_{\rm NL} &= 0 \; {\rm [h}^{-1}{\rm Mpc]}^{2}: \; \alpha_{\rm cross} = 0.985 \pm 0.057 \nonumber \\
\Sigma^{2}_{\rm NL} &= 6 \; {\rm [h}^{-1}{\rm Mpc]}^{2}: \; \alpha_{\rm cross} = 0.982 \pm 0.058 \nonumber
\end{eqnarray}
}

\begin{figure}
	\centering
    \includegraphics[width=\columnwidth]{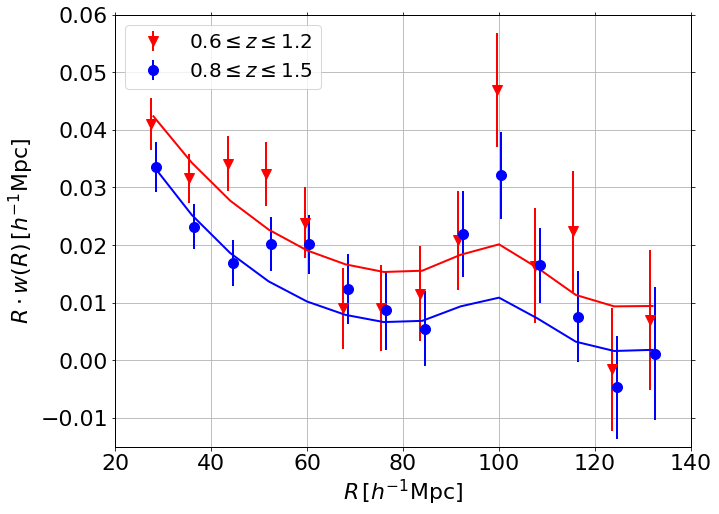}
    \vskip -0.2cm
    \caption{The projected cross-correlation function for $0.6 \leq z \leq 1.2$ (red) and $0.8 \leq z \leq 1.5$ (blue) and the acoustic scale region. The solid curves represent the best-fitting model in each case.}
    \label{fig:bao-cross} 
\end{figure}

Fig.~\ref{fig:bao-cross} displays the BAO measurements for the weighted mean of the projected cross-correlation function when using quasars in $0.6 \leq z \leq 1.2$ (red) and $0.8 \leq z \leq 1.5$ (blue) together with the best-fitting model in solid curves for our fiducial configuration: bin width of 8$h^{-1}$Mpc, fitting range $20 < s < 140h^{-1}$Mpc and a covariance matrix from 100 jackknife realisations. We can see that the amplitude of the cross-correlation signal is higher for the redshift range $0.6 \leq z \leq 1.2$, meaning that more photometric ELGs lie in this redshift range as expected according to the photometric redshift distribution shown in Fig.~\ref{fig:photo-z}.

Table~\ref{tab:BAO-cross} summarises the results of the BAO fits obtained from the projected cross-correlation in the two redshift ranges we consider for our fiducial configuration with a bin width of 8~$h^{-1}$Mpc, a fitting range $20 < s < 140h^{-1}$Mpc, a covariance matrix from 100 jackknife realisations. \com{Despite the apparent bad fits, the $\chi^{2}$ are fine and we remind that the data points are very correlated with each other.} As for the auto-correlation function, we also fit the individual jackknife realisations and report the mean result for both the value and the error bar. As expected, both the full sample and the mean of the jackknife regions are consistent with each other.
Table~\ref{tab:BAO-cross} also shows the results of the consistency tests \com{when including the damping term $\Sigma_{\rm NL}$ which accounts for non-linear effects}, when varying the binning, the fitting range and when using different imaging weights for the photometric sample. \com{We also find very similar results on $\alpha_{\rm cross}$ between the three definitions of the effective redshift described in Section~\ref{sec:effective-z}.} All the results are consistent with each other within 1$\sigma$, demonstrating the robustness of the BAO feature in the projected cross-correlation function. 

\begin{table}
    \caption{Results for the projected BAO fits to the cross-correlation of the DR16 eBOSS quasars with ELG galaxies from DESI DR8 Legacy Imaging Surveys. The fiducial configurations uses data with 8$h^{-1}$Mpc bin size and centres in the range $10 < s < 140h^{-1}$Mpc and a covariance matrix from 100 jackknife realisations.}
    \label{tab:BAO-cross}
    \begin{tabular}{|c|c|c|}
    \hline
    Configuration & $\alpha_{\rm cross}$ & $\chi^{2}$/d.o.f. \\
    \hline
    \multicolumn{3}{c}{$0.8 \leq z \leq 1.5$} \\
    \hline
    Fiducial                    & 0.994 $\pm$ 0.051 & 9.1/9 \\
    mean of the jackknifes                    & 0.994 $\pm$ 0.051 & 9.3/9 \\
    $\Sigma_{\rm NL}$ = 6 [h$^{-1}$Mpc]$^{2}$ & 0.993 $\pm$ 0.055 & 9.4/9 \\
    $\Delta s$ = 5$h^{-1}$Mpc   & 1.005 $\pm$ 0.047 & 17.3/19 \\
    $20 < s < 150h^{-1}$Mpc     & 0.997 $\pm$ 0.049 & 9.4/11 \\
    $10 < s < 140h^{-1}$Mpc     & 0.998 $\pm$ 0.052 & 10.8/11 \\
    no $w_{\rm sys}$            & 0.995 $\pm$ 0.044 & 8.6/9 \\
    $w_{\rm sys-lin}$           & 0.989 $\pm$ 0.046 & 7.5/9 \\
    $w_{\rm sys-quad}$          & 0.988 $\pm$ 0.046 & 8.4/9 \\
    $w_{\rm sys-nn-fs}$         & 0.993 $\pm$ 0.046 & 9.4/9 \\
    NGC                         & 0.970 $\pm$ 0.066 & 7.8/9 \\    
    SGC                         & 1.025 $\pm$ 0.109 & 5.9/9 \\ 
    \hline
    \multicolumn{3}{c}{$0.6 \leq z \leq 1.2$} \\
    \hline
    Fiducial                    & 0.999 $\pm$ 0.059 & 13.1/9 \\
    mean of the jackknifes                   & 0.999 $\pm$ 0.059 & 13.3/9 \\
    $\Sigma_{\rm NL}$ = 6 [h$^{-1}$Mpc]$^{2}$ & 0.998 $\pm$ 0.063 & 13.2/9 \\
    $\Delta_{s}$ = 5$h^{-1}$Mpc & 1.014 $\pm$ 0.058 & 20.2/19 \\
    $20 < s < 150h^{-1}$Mpc     & 0.991 $\pm$ 0.059 & 14.8/11 \\
    $10 < s < 140h^{-1}$Mpc     & 1.003 $\pm$ 0.061 & 13.5/11 \\
    no $w_{\rm sys}$            & 1.004 $\pm$ 0.057 & 11.3/9 \\
    $w_{\rm sys-lin}$           & 0.997 $\pm$ 0.059 & 12.9/9 \\
    $w_{\rm sys-quad}$          & 0.994 $\pm$ 0.058 & 13.7/9 \\
    $w_{\rm sys-nn-fs}$         & 0.999 $\pm$ 0.056 & 13.4/9 \\
    NGC                         & 0.965 $\pm$ 0.078 & 16.2/11 \\    
    SGC                         & 1.045 $\pm$ 0.112 & 5.3/11 \\  
    \hline
    \end{tabular}
\end{table} 



\subsection{BAO constraints and discussion}

\begin{figure}
	\centering
    \includegraphics[width=0.8\columnwidth]{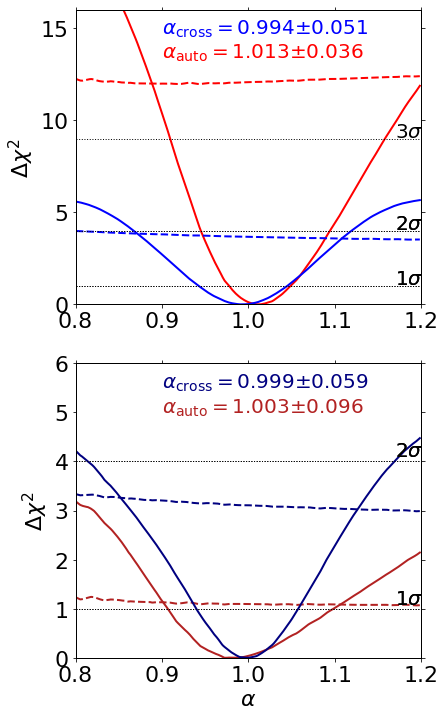}
    \vskip -0.2cm
    \caption{Likelihood of the transverse BAO parameter $\alpha_{\rm cross}$ in terms of $\Delta \chi^{2}$ in the redshift ranges $0.8 \leq z \leq 1.5$ (top) and $0.6 \leq z \leq 1.2$ (bottom). In each panel, the solid curves display the likelihood obtained when fitting the data with a model that contains the BAO feature, while the dashed curves display the same information for a model without BAO.}
    \label{fig:bao-significance} 
\end{figure}

Fig.~\ref{fig:bao-significance} displays the likelihood and BAO detection significance in terms of $\Delta \chi^{2}$ obtained from the auto- and projected cross-correlation function in $0.8 \leq z \leq 1.5$ (top) and $0.6 \leq z \leq 1.2$ (bottom). The dashed curves represent the template without the BAO feature in each case. The likelihoods from the auto-correlation function (red and dark red) are more skewed towards large values of $\alpha$ compared to the ones from the projected cross-correlation function. In both redshift ranges, the BAO detection significance is about 2$\sigma$ for the projected cross-correlation function while it is greater than 3$\sigma$ in $0.8 \leq z \leq 1.5$ and less than 1.5$\sigma$ in $0.8 \leq z \leq 1.5$ for the auto-correlation function. 

In both redshift ranges, the statistics are too low to enable an anisotropic BAO measurement from the auto-correlation function of the eBOSS DR16 quasar sample but \citet{Hou+20,Neveux+20} did the anisotropic BAO fitting using eBOSS DR16 quasars in $0.8 \leq z \leq 2.2$. 
Both techniques provide consistent results within less than 0.5$\sigma$, showing the robustness of the BAO feature in galaxy clustering. However, the BAO shifts are not sensitive to exactly the same cosmic distance as presented in Section~\ref{sec:parameter}. The BAO shift from the auto-correlation is sensitive to the spherically-averaged distance $D_{\rm V}$ while the BAO shift from the projected cross-correlation function is sensitive to the angular diameter distance $D_{\rm M}$ such that we obtain:
\begin{eqnarray}
0.8 \leq z \leq 1.5 \; z_{\rm eff}=1.20 \nonumber \\
    \rm auto: D_{\rm V}(z_{\rm eff})/r_{\rm drag}&=&26.53\pm{0.94} \nonumber \\
    \rm cross: D_{\rm M}(z_{\rm eff})/r_{\rm drag}&=&30.42\pm{1.56} \nonumber \\
0.6 \leq z \leq 1.2 \; z_{\rm eff}=0.92 \nonumber \\
    \rm auto: D_{\rm V}(z_{\rm eff})/r_{\rm drag}&=&26.3\pm{2.5} \nonumber \\
    \rm cross: D_{\rm M}(z_{\rm eff})/r_{\rm drag}&=&30.6\pm{1.8} \nonumber
\end{eqnarray}

In the redshift range $0.6 \leq z \leq 1.2$ where we expect to have more ELG, we indeed obtain a more precise distance measurement with the projected cross-correlation function than with the auto-correlation function (5.9\% against 9.5\%). However, as mentioned above the BAO detection significance remains low in both cases. 
The comparison between the two redshift ranges suggests that it is essential to ensure the best overlap in redshift between the spectroscopic and photometric samples. Moreover, it seems to suggest that although the spectroscopic target density may be too low to obtain a strong BAO detection in the auto-correlation function, we can expect a stronger detection in the projected cross-correlation function. Table~\ref{tab:summary} summarises the configuration for each redshift range with the target density for each sample and the precision on the BAO scale obtained from the auto- and projected cross-correlation function. The first two rows correspond to this work where the galaxy density of the photometric sample in each redshift range is computed using the photometric redshifts of Fig.~\ref{fig:photo-z}. We also show the configuration in~\citet{Patej+18} in the third row but the authors did not provide a BAO measurement from the auto-correlation function of their spectroscopic sample, this is why we did not quote a precision for $\sigma_{\rm auto}$ in $0.6 < z < 0.8$. The number densities in our analysis are more optimal for this type of cross-correlation (mainly a denser photometric sample) so  we may have expected a more precise measurement from the projected-cross correlation function but we also have more important systematics in the photometric sample obtained from the DESI Legacy Imaging Surveys as we pushed towards very faint objects to reach a high sampling of galaxies at high redshifts ($z > 1$). Moreover, \citet{Patej+18} used a narrower redshift bin which could also help improve the constraint from the projected cross-correlation function.

\begin{figure}
	\includegraphics[width=\columnwidth]{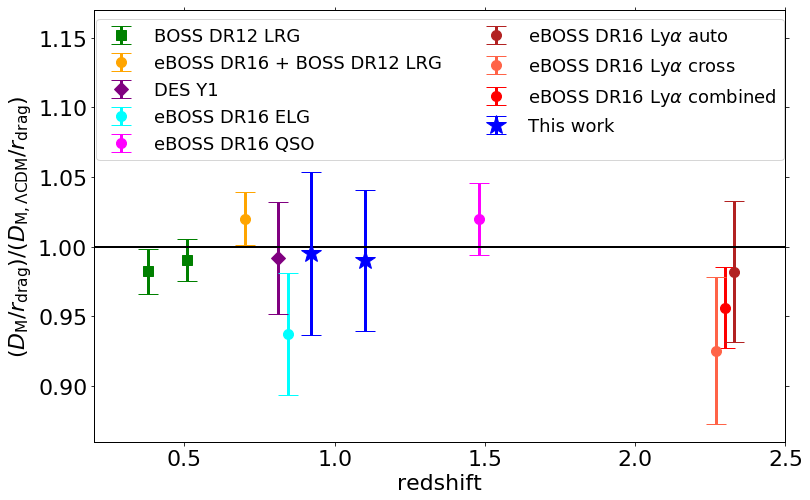}
    \vskip -0.2cm
    \caption{Measurements of the comoving angular diameter distance $D_{\rm M}$ as a function of redshift. Our measurements at $z_{\rm eff}=0.92$ and $z_{\rm eff}=1.1$ are shown using a blue star.}
    \label{fig:baoplot} 
\end{figure}

However, the comparison with the auto-correlation function is very encouraging, showing that not only we can detect the BAO feature in the projected cross-correlation function but also put better constraints when the spectroscopic target density is too low to enable a strong BAO detection in the auto-correlation function. We also compare our measurements of $D_{\rm M}$ with other measurements using different tracers and methods, as shown in Fig.~\ref{fig:baoplot}: BOSS DR12 LRG results using BAO only~\citep{boss-DR12}; DES Y1 using photometric data in a combined analysis with weak lensing and clustering~\citep{DES17}; the eBOSS final results
~\citep{eBOSS20} including eBOSS+BOSS LRG using BAO only~\citep{Bautista+20,GilMarin+20}, eBOSS ELG using BAO+RSD~\footnote{The statistics of the eBOSS ELG sample is not enough to perform an anisotropic BAO fitting and thus to measure $D_{\rm M}$ from BAO only.}~\citep{deMattia+20, Tamone+20}, eBOSS QSO using BAO only~\citep{Hou+20,Neveux+20} and eBOSS Ly-$\alpha$ forests~\citep{duMasDesBourboux+20} using BAO from the auto-correlation Ly-$\alpha$/Ly-$\alpha$ (dark red), cross-correlation Ly-$\alpha$/quasar (light red) and the combined measurement (red). This latest example shows the gain in precision on the combined measurement from auto- and cross-correlation. Although the analysis in this work is very different from the one of the Ly-$\alpha$ forests, in future work it would be interesting to investigate the potential gain on $D_{\rm M}$ of combining the auto- and projected cross-correlation function.

\begin{table*}
    \caption{Comparison between the auto-correlation and the projected cross-correlation functions for this work (top rows) and \citet{Patej+18} (bottom row) in terms of redshift range, target density and constraints on the BAO scale.}
    \label{tab:summary}
    \begin{tabular}{|c|c|c|c|c|c|}
    \hline
    Redshift range & area [deg$^{2}$] & spectro [deg$^{-2}$] & photo [deg$^{-2}$] & $\sigma_{\rm auto}$ & $\sigma_{\rm cross}$ \\
    \hline
    $0.8 \leq z \leq 1.5$ & 4000 & 35 & 2100 & 3.5\% & 5\%\\
    $0.6 \leq z \leq 1.2$ & 4000 & 20 & 2900 & 9\% & 6\% \\
    \hline
    $0.6 < z < 0.8$ & 6000 & 35 & 1100 & -- & 3\% \\
    \hline
    \end{tabular}
\end{table*}

\section{Conclusions}
\label{sec:concl}
We have applied a method proposed in~\citet{Patej+18} based on the cross-correlation between a sparse spectroscopic sample and a denser photometric sample to constrain the angular diameter distance by searching for the transverse BAO. We have used a sample of SDSS-IV eBOSS quasars between $0.6 \leq z \leq z 1.5$ and we have produced a high density sample of galaxies using the DESI Legacy Imaging Surveys. Since we need to select fainter objects at the limit of the survey depth, we expect the photometric sample to be more prone to density fluctuations due to inhomogeneities in the selection. To mitigate for this effect, we have applied the neural network technique developed in~\citet{Rezaie+19} to our photometric sample and confirmed that it can correct for complex variations that standard multivariate linear regression techniques cannot. We have validated the pipeline of the projected cross-correlation function against approximate mocks and we have demonstrated that the BAO measurement with this method is robust against a variety of observational choices.

We have performed two analyses in parallel: the auto-correlation of the eBOSS quasars and the projected cross-correlation function in order to provide a detailed comparison.
We have investigated two configurations: one where we cross-correlate the photometric galaxies with quasars in $0.8 \leq z \leq 1.5$ and another one with quasars in $0.6 \leq z \leq 1.2$. In the latter, we find that the cross-correlation technique can reduce shot noise and thus provide better constraints on the cosmic distance (6\% precision) compared to the results obtained from the auto-correlation (9\% precision). However, we also find that we are limited by the number density and purity of the photometric sample and its overlap in redshift with the spectroscopic sample, which thus affects the performance of the method.
We also find a noticeable peak in the acoustic scale region which is larger than expected in usual theories, although the fits with templates based on the matter power spectrum yield only a 2$\sigma$ indication. We compare the amplitude of the signal in the acoustic scale region with approximate mocks and we perform a series of consistency tests that indicate no bias on the cosmological constraint we derive from the fits of the projected cross-correlation function. Nevertheless, we highlight that DESI will soon start its cosmological survey and it would be worth checking whether the discrepancy at the BAO scale remains when the clustering catalogues are available.

In addition, the technique will be even more promising with the arrival of deeper photometric data thanks to upcoming surveys such as \textit{Euclid}~\citep{Euclid13}. The method could be applied to DESI quasars, which will still be limited by shot noise~\citep{Desi16a}, with a sample of H$\alpha$ ($0.7 < z < 2$ and [OIII] emission galaxies ($2 < z < 2.7$) selected from \textit{Euclid}~\citep{Mehta+15}. This should enable to put better constraints on the transverse BAO scale at $z \geq 2$ than DESI will with the auto-correlation function of the quasars alone.

\section*{Acknowledgements}
PZ would like to thank John Helly for his help modifying TWOPCF.
PZ, SC and PN acknowledge support from the Science Technology Facilities Council through ST/P000541/1 and ST/T000244/1.
M.R. is supported by the U.S.~Department of Energy, Office of Science, Office of High Energy Physics under DE-SC0014329; H.-J.S. is supported by the U.S.~Department of Energy, Office of Science, Office of High Energy Physics under DE-SC0014329 and DE-SC0019091.
SA is supported by the European Research Council through the COSFORM Research Grant (\#670193).
This work used the DiRAC@Durham facility managed by the Institute for Computational Cosmology on behalf of the STFC DiRAC HPC Facility (www.dirac.ac.uk). The equipment was funded by BEIS capital funding via STFC capital grants ST/K00042X/1, ST/P002293/1 and ST/R002371/1, Durham University and STFC operations grant ST/R000832/1. DiRAC is part of the National e-Infrastructure. Funding for the Sloan Digital Sky Survey IV has been provided by the Alfred P. Sloan Foundation, the U.S. Department of Energy Office of Science, and the Participating Institutions. SDSS-IV acknowledges
support and resources from the Center for High-Performance Computing at
the University of Utah. The SDSS web site is www.sdss.org.
SDSS-IV is managed by the Astrophysical Research Consortium for the 
Participating Institutions of the SDSS Collaboration including the 
Brazilian Participation Group, the Carnegie Institution for Science, 
Carnegie Mellon University, the Chilean Participation Group, the French Participation Group, Harvard-Smithsonian Center for Astrophysics, 
Instituto de Astrof\'isica de Canarias, The Johns Hopkins University, Kavli Institute for the Physics and Mathematics of the Universe (IPMU) / 
University of Tokyo, the Korean Participation Group, Lawrence Berkeley National Laboratory, 
Leibniz Institut f\"ur Astrophysik Potsdam (AIP),  
Max-Planck-Institut f\"ur Astronomie (MPIA Heidelberg), 
Max-Planck-Institut f\"ur Astrophysik (MPA Garching), 
Max-Planck-Institut f\"ur Extraterrestrische Physik (MPE), 
National Astronomical Observatories of China, New Mexico State University, 
New York University, University of Notre Dame, 
Observat\'ario Nacional / MCTI, The Ohio State University, 
Pennsylvania State University, Shanghai Astronomical Observatory, 
United Kingdom Participation Group,
Universidad Nacional Aut\'onoma de M\'exico, University of Arizona, 
University of Colorado Boulder, University of Oxford, University of Portsmouth, 
University of Utah, University of Virginia, University of Washington, University of Wisconsin, 
Vanderbilt University, and Yale University.
In addition, this research relied on resources provided to the eBOSS
Collaboration by the National Energy Research Scientific Computing
Center (NERSC).  NERSC is a U.S. Department of Energy Office of Science
User Facility operated under Contract No. DE-AC02-05CH11231.
This research is supported by the Director, Office of Science, Office of High Energy Physics of the U.S. Department of Energy under Contract No., and by the National Energy Research Scientific Computing Center, a DOE Office of Science User Facility under the same contract; additional support for DESI is provided by the U.S.
National Science Foundation, Division of Astronomical Sciences under Contract No. AST-0950945 to the National Optical Astronomy Observatory; the Science and Technologies
Facilities Council of the United Kingdom; the Gordon and
Betty Moore Foundation; the Heising-Simons Foundation; the French Alternative Energies and Atomic Energy Commission (CEA); the National Council of Science and Technology of Mexico, and by the DESI Member Institutions.
The Legacy Imaging Surveys of the DESI footprint are supported by the Director, Office of Science, Office of High Energy Physics of the U.S. Department of Energy under Contract No. DE-AC02-05CH1123, by the National Energy Research Scientific Computing Center, a DOE Office of Science User Facility under the same contract; and by the U.S. National Science Foundation, Division of Astronomical Sciences under Contract No. AST-0950945 to NOAO.

\section*{Data availability}

The photometric data underlying this article are publically available as part of the DESI Legacy Imaging Surveys at \url{https://www.legacysurvey.org/dr8/description/}.
The spectroscopic data underlying this article are available to the public at
\url{https://data.sdss.org/sas/dr16/eboss/lss/catalogs/DR16/} and are described at \url{https://www.sdss.org/dr16/spectro/lss/}.

\appendix

\section{Neural network with feature selection}
\label{sec:appendixA}

\begin{figure}
	\centering
    \includegraphics[width=\columnwidth]{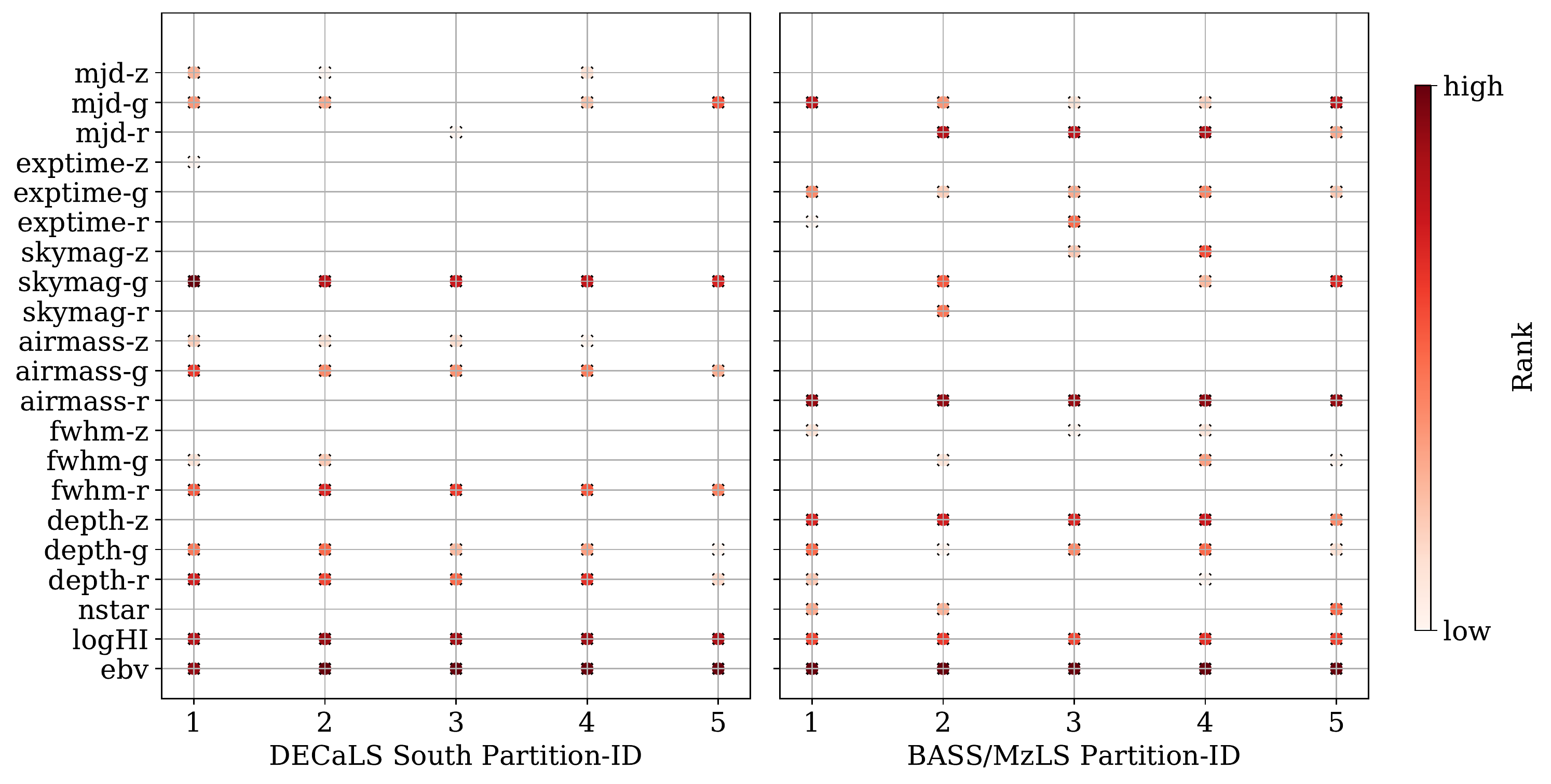}
    \caption{Important imaging maps according to the feature selection procedure for DECaLS-South (left) and BASS/MzLS (right). \com{The selected maps are shown with circles and their rank is colour-coded to represent the more important maps with the darker circles.}}
    \label{fig:NN-fs-maps} 
\end{figure}

Because the imaging attributes are correlated, they can contain redundant information which increases the risk of over-fitting and degrading the cosmological clustering. Within the neural network framework, we can apply a feature selection process to identify the redundant and irrelevant imaging maps (and thus reduce the number of input imaging attributes) by splitting the data into 5 partitions. We train a linear model on all the input imaging quantities and then eliminate one and train again the model on the remaining input features. Note that given the definition of the minimisation function, if the feature contains relevant information on the systematic effect, it would make the fit worse \com{ and the model would yield a higher validation error}. \com{This procedure removes one map at a time iteratively, and finally it ranks the imaging maps such that} the features which produce the highest improvement in fitting are removed. Fig.~\ref{fig:NN-fs-maps} shows the results of the feature selection procedure described above where the darker the dot, the more important the imaging attribute, for DECaLS-South (left) and BASS/MzLS (right).
We can see that the most important imaging systematics are not the same depending on the region, which confirms the importance of treating the different surveys separately. In both regions, Galactic extinction (ebv) and the hydrogen atom column density (logHI) are important, such Galactic depths, airmass and sky brightness but in different bands depending on the region.

\begin{figure}
	\centering
    \includegraphics[width=\columnwidth]{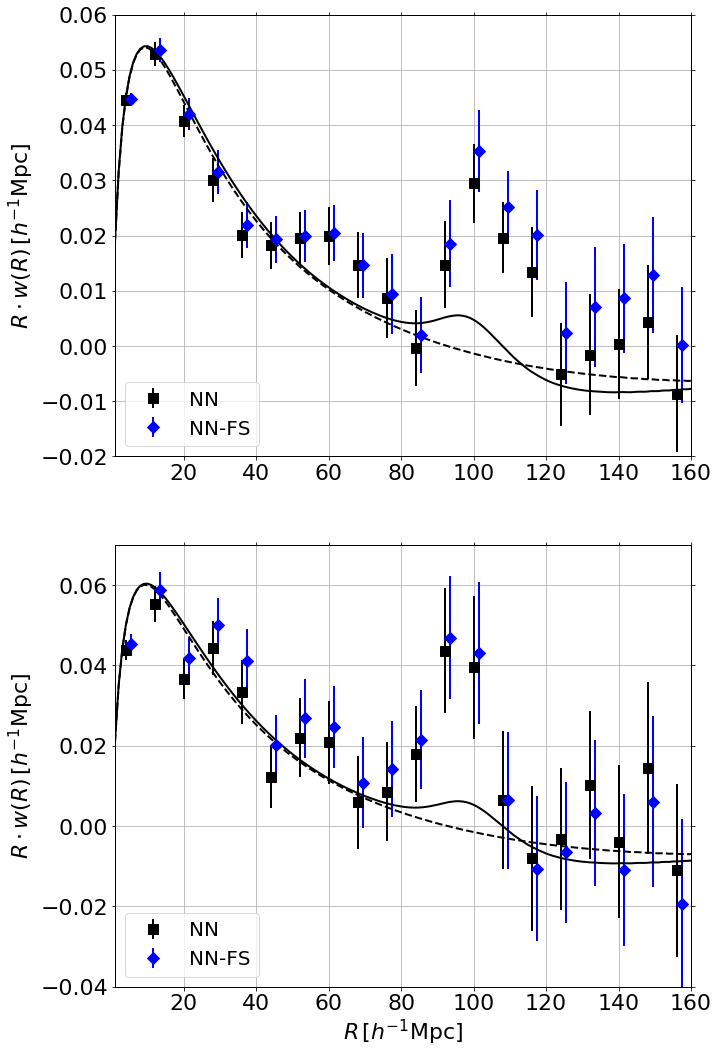}
    \vskip -0.2cm
    \caption{Projected cross-correlation function in $0.8 \leq z \leq 1.5$ in the BASS/MzLS (top) and DECaLS-South (bottom) regions after applying neural network with feature selection (blue) and without (black, our baseline).}
    \label{fig:NN-fs-baocross} 
\end{figure}

After identifying the most important imaging attributes, we derived a new set of photometric weights that account for the feature selection. In Fig.~\ref{fig:NN-fs-baocross}, we compare the projected cross-correlation function after applying this new set of neural network weights (NN-FS) with our baseline where the feature selection was not applied. The top panel corresponds to BASS/MzLS in the eBOSS footprint and the bottom panel to DECaLS-South, for the redshift range $0.8 \leq z \leq 1.5$. The effect of including the feature selection is marginal and we found the same behaviour for $0.6 \leq z \leq 1.2$. For this reason, we kept the baseline to be the one without feature selection. We also check the consistency in terms of BAO constraints and the results when including the feature selection is shown in Table~\ref{tab:BAO-cross}.

\section{Impact of mixing different imaging surveys}
\label{sec:appendixB}

In what follows, in order to avoid mixing the surveys and because the analysis is limited to the eBOSS footprint, we remove the region at dec $< 32.375$ deg in both the spectroscopic and photometric samples to consider BASS-MzLS only in the NGC. We also remove the DES region in the SGC (dec $< 5$ deg) as it is deeper than DECaLS. Given that we restrict to the eBOSS footprint, the removed regions have small areas so the statistical precision is only slightly affected but in this section, we show that mixing these surveys degrades the cosmological signal significantly.

\begin{figure}
	\centering
    \includegraphics[width=\columnwidth]{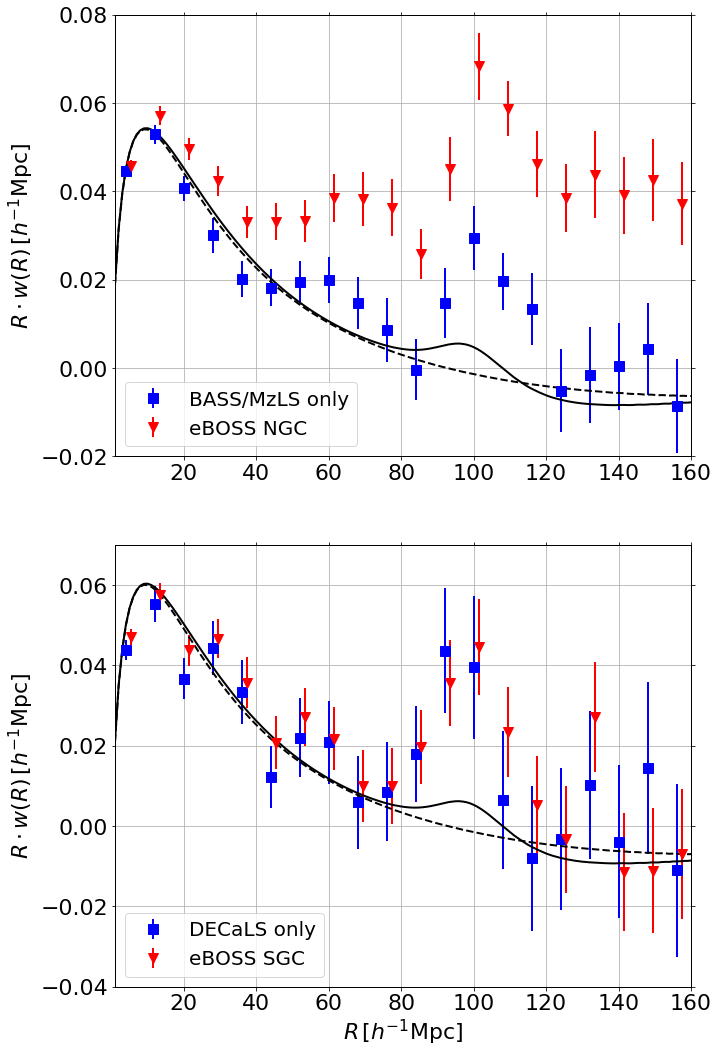}
    \vskip -0.2cm
    \caption{Impact of removing DECaLS-North in the eBOSS NGC (top) and DES in the eBOSS SGC (bottom) on the projected cross-correlation function. The measurements used in the main analysis correspond to the ones in blue. The cosmological signal is more degraded in the NGC as we combine different imaging surveys using different observing sites (BASS/Mzls and DECaLS) while in the SGC, we are using the same imaging camera but DES goes deeper than DECaLS.}
    \label{fig:appendixB1} 
\end{figure}

In Section~\ref{sec:photo-data}, we describe the different imaging surveys we use to select the photometric sample and we show that the observed galaxy density correlates with imaging systematics differently depending on the survey. When we restrict to the eBOSS footprint, the NGC contains both BASS/MzLS at dec $>$ 32.375 deg and DECaLS-North below while the SGC contains both DECaLS and DES at dec $<$ 5 deg. In the main analysis, we decide to remove DECaLS-North in the NGC and DES in the SGC.
Fig.\ref{fig:appendixB1} displays the impact on the projected cross-correlation function of removing those regions in the NGC (top) and in the SGC (bottom). \com{In all cases, we show the projected cross-correlation function after applying the NN weights. The measurements in blue correspond to the ones used in the main analysis when considering BASS/MzLS only in the NGC and DECaLS-S only in the SGC.} We can see that the signal is largely improved when we consider BASS/MzLS only in the NGC. The effect of removing DES is less pronounced.

\begin{figure}
	\centering
    \includegraphics[width=\columnwidth]{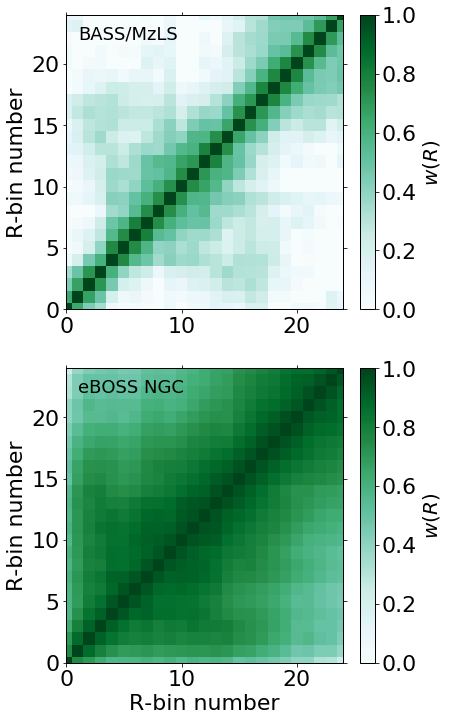}
    \vskip -0.2cm
    \caption{Correlation matrix of the projected cross-correlation function obtained from the 100 jackknife regions for BASS/MzLS only in the eBOSS NGC footprint (top) and for the entire eBOSS NGC (bottom) with two different imaging surveys, BASS/MzLS and DECaLS.}
    \label{fig:appendixB2} 
\end{figure}

Fig.\ref{fig:appendixB2} displays the correlation matrix of the projected cross-correlation function obtained from the 100 jackknife regions for BASS/MzLS only (top) and for the entire eBOSS NGC (bottom). \com{It seems that mixing the different surveys in the North brings inhomogeneities in the photometry which degrade the cosmological signal. Treating the DESI imaging surveys separately is therefore essential to preserve the cosmological information. Moreover, given that the main DESI ELG selection is also pushed towards faint magnitudes, there is an important ongoing effort within the DESI collaboration to study the impact of heterogeneous photometry on the clustering of these faint galaxies. The DESI ELG EZ mocks presented in Section~\ref{sec:clust-cross} have been developed with this purpose in mind, in order to perform BAO and Full-Shape analyses on these mocks and quantify the effect on the cosmological parameters.}
 
\bibliographystyle{mnras}
\bibliography{main} 

\appendix


\bsp	
\label{lastpage}
\end{document}